\documentclass[
nofootinbib,
 amsmath,amssymb,
 aps,
superscriptaddress]{revtex4-2}

\usepackage{graphicx}
 \usepackage{float}
\usepackage{dcolumn}
\usepackage{bm}

\usepackage[usenames,dvipsnames]{xcolor}
\usepackage{color} 
\usepackage[linktocpage,breaklinks]{hyperref}
\hypersetup{
    colorlinks=true, 
    pdfborder = {0 0 0.5 [3 3]},
    anchorcolor=black,
    citecolor=NavyBlue, 
    linktoc=all,    
    linktocpage=true,
    linkcolor=NavyBlue,
	urlcolor=NavyBlue, 
}

\usepackage{tikz}
\usetikzlibrary{mindmap,shadows,positioning}
\definecolor{myTeal}{HTML}{A4C8C8}
\def\Sc{Schwarzschild }

\begin{document}

\title{Black-Hole mimickers in GR and $f(R)$ gravity}




\author{Hodek M. García}
\email{hodek.mealstrom@ciencias.unam.mx}
\address{Instituto de Ciencias Nucleares, Universidad Nacional Autónoma de México, A.P. 70-543, México CDMX. 04510, México}
 \author{Marcelo Salgado}%
 \email{marcelo@nucleares.unam.mx}
\address{Instituto de Ciencias Nucleares, Universidad Nacional Autónoma de México, A.P. 70-543, México CDMX. 04510, México} 
\address{Departament  de  F\'isica,  Universitat  de  les  Illes  Balears,  Palma  de  Mallorca,  E-07122,  Spain}





\date{\today}

\begin{abstract}
Black hole mimickers (BHMs) are horizonless globally regular ultracompact relativistic self-gravitating objects (UCOs) of mass $M$ and radius $R$ with compactness ${\cal C} = M/R$ higher than that of a neutron star and that produce an effective 
potential for null geodesics (photons) that possesses a local maximum, which is usually accompanied by an inner local minimum. The presence of a local maximum allows for unstable circular orbits to exist similar to light rings present in actual BH solutions, while it has been conjectured that the presence of a local minimum is symptomatic 
of potential instabilities. One such candidate for a BHM is a solitonic boson star (SBS) which 
is a boson star endowed with a sextic potential. In this paper we investigate further 
solutions of static and spherically symmetric SBSs in general relativity with a larger set of parameter values, and argue that such solutions are very similar to UCOs composed of an incompressible perfect fluid (IPF) with a sufficiently  large pressure (the mimicker of a BHM). These IPFUCOs reach the Buchdahl limit ${\cal C}= 4/9$ for 
arbitrarily large pressures. We investigate the extent to which the IPFUCOs constructed within a quadratic model in $f(R)$ gravity can overcome this limit or not, and thus pave the way for possibly building SBSs (or other kind of UCO) within this (or other alternative theory of gravity). We further elaborate about the stability properties of SBSs which have been the subject of some controversy recently.


\end{abstract}



\maketitle


\section{Introduction}

Black holes (BHs) are some of the most enigmatic {\it objects} in the universe, 
and perhaps the most astonishing prediction of Einstein's general relativity 
(GR) (see \cite{Wald,HawkingEllis,Chandrasekhar} for a thorough review). The BH's fingerprint is the {\it event horizon} (EH), a three dimensional null surface
(with topology $\mathbb{R}\times S^2$) that separates its interior from the exterior (the observable universe) beyond which any signal cannot escape to the exterior (the domain of outer communication or DOC). 
An extended object, like a spherical star, that collapses gravitationally and whose matter enters into its \Sc radius 
forms a BH. A static observer near the EH but outside the BH has a proper time that 
seems to freeze when compared with the time measured by a far-away observer (FAO) from 
the BH (this is due to gravitational time dilation). Owing to this feature, a light signal emitted near the EH but in the DOC 
is highly red-shifted when detected by the FAO. Moreover, when light signals are sent 
towards the BH with a low impact parameter and high energy can be captured by the light sphere located outside the EH 
(at $r=3M$\footnote{In units where $G=c=1$} in the \Sc BH scenario) which is associated with {\it unstable} circular 
orbits. Photons can remain there while turning around the BH before being scattered to infinity or swallowed by the BH
leading to light ring (LR) images when detected by a FAO. This is
an example of an extreme light bending and the one responsible for producing multiple images when the light
emitted from astrophysical sources from the background passes near the photon sphere of a BH before getting scattered and detected by a FAO. A prominent example of this phenomenon is when a hot accretion disk of gas revolves around a BH and the light emitted from this disk passes near the photon sphere, producing multiple LRs associated with the different parts of the disk. This scenario has been confirmed by the Event Horizon Telescope (EHT) when looking at 
the centers of some galaxies (M81 and the Milky Way)  \cite{EHT1,EHT2}. These images can suffer further (red and blue shift) distortions if in addition the BH rotates with respect the FAO, which is the case of a Kerr BH.

BHs are also remarkable since they seem to have properties analogous to thermodynamic systems, namely, the 
so called {\it laws of BH thermodynamics} \cite{Wald}. Taking into account the quantum aspects of fields 
gravitating around BHs, Hawking has shown that these thermodynamic analogues are more than a coincidence and that 
BHs can emit thermal black-body radiation with a temperature proportional to their surface gravity 
\cite{Hawkingrad}. 

One of the most intriguing aspects of BHs is that their classical 
(as opposed to quantum) description entails the possible appearance of physical 
{\it singularities} in its interior when the matter, during a gravitational collapse of a star, 
starts to clump around its center with higher and higher densities as it continues to fall towards the interior of the star. These singularities represent a breakdown of 
GR where curvature and density become infinite. Penrose's singularity theorems (see \cite{Wald,HawkingEllis} for a 
review) indicate that such singularities are an unavoidable fate of gravitational collapse under generic circumstances (e.g. generic initial conditions) and when matter obeys some energy conditions (e.g. the strong energy condition). When these assumptions hold in GR BH formation becomes a paramount feature during gravitational collapse as those singularities remain hidden from the DOC behind the EH protecting the 
observable universe from their eerie and fatal consequences (e.g. radiation of infinite energy). This is the famous {\it cosmic censorship conjecture} (e.g. see \cite{Wald})
that establishes that BHs always form prior the onset of singularities at the interior of a collapsing star. 

It is widely believed that a quantum theory of gravity (QTG) should cure the classical theory from the formation of such singularities, much like QED (or simply QM) 
has cured the singularities associated with classical charged point particles. Loop 
quantum gravity, for instance, is an example of a QTG that provides such a cure, 
at least in simplified scenarios of gravitational collapse \cite{Rovelli}. However, it is 
unclear if in more realistic situations this cure is always available. But regardless 
of this undesirable aspect of classical GR, one believes that the quantum effects 
develop after the BH has formed and when the density of the infalling matter 
is close to the Planck's density. Notwithstanding, BH formation is considered to 
be conceptually independent from the emergence or not of singularities. 
In fact, a long-standing question within GR is if one can find BH solutions 
that are free of singularities. Under specific assumptions, exotic matter-field models do yield an affirmative answer, like non-linear electrodynamics or in models of modified gravity 
(see \cite{Huang2025} for an updated discussion). Nevertheless, 
it is unclear if such regular BHs can form dynamically from suitable and generic initial conditions.

In addition to all these traits, BHs are also the source of 
some paradoxes, like the infamous {\it loss of information paradox} \cite{lossinfo}, which is related with Hawking's 
BH evaporation process alluded above and remains an
unresolved problem of theoretical physics  (or at least no universal consensus exists 
among the physics community about its resolution) \cite{Okon,Maudlin}.

Two of the best-known BH solutions in GR are the aforementioned \Sc and Kerr 
solutions, which are solutions of the Einstein field equations in vacuum and for asymptotically flat (AF) spacetimes. 
These are contained within
the Kerr-Newman (KN) three-parameter family, a solution in GR in the presence of electromagnetic fields. 
For instance, the Reissner–Nordström (RN) solution is recovered from this family in the absence of rotation, while the 
\Sc and Kerr solutions are recovered in the absence of electromagnetic charges.

Another undesirable property of some BH solutions (e.g. RN and Kerr) is the presence of 
{\it Cauchy horizons}, null surfaces beyond which the spacetime cannot be predicted from initial data provided on a spacelike hypersurface (e.g. see \cite{Wald,HawkingEllis}), although this data 
are indeed able to predict its causal future in the DOC. These horizons, which are located inside the BH, might be just a mathematical curiosity arising from the analytic extensions of BH solutions to their interior, but which could be unstable when introducing spacetime and matter perturbations, as well as quantum effects \cite{Dafermos2003,Hollands2020}.

From the observational point of view, there is strong evidence that BHs not only 
exist in nature, but most, if not all of them, are (or appear to be) of Kerr or \Sc type from an exterior observer. 
This evidence comes from multiple and independent observations. For instance, 
the detection of gravitational waves (GW) by the LIGO-Virgo-KAGRA collaboration \cite{LVK} together with numerical simulations \cite{Ninja,Ninja2} indicate that these waves are the result of the inspiraling of two 
BHs (Kerr or Schwarzschild) that merge into a single BH that settles into a stationary Kerr BH after 
the ringdown phase of the merger. In a few instances so far these signals come from the merger of two neutron stars (NS). 
Nevertheless, in this case, GWs are also accompanied by electromagnetic and neutrino counterparts, which allow them to differentiate from a binary BH collision. Moreover, the binary NS scenario cannot account for the existence of progenitors with tenths or more solar masses, which 
are required to produce many of the observed GW signals.

Another evidence comes from the images taken by the EHT \cite{EHT1,EHT2} which are compatible with the Kerr hypothesis. Finally, the dynamics of stars around the Milky Way is also compatible with the BH hypothesis at the center 
of our galaxy \cite{Ghez1,Ghez2,Ghez3,Ghez4,Genzel,Abuter}.

While the BH evidence in the universe is overwhelming, albeit indirect so far, it is possible that other alternatives to BH 
may exist. Some of these alternatives have been put forward mainly based on theoretical and epistemological arguments with different motivations in mind. As argued above, one 
motivation is to find solutions in GR (or in other alternative theory 
of gravity) free of singularities and of other undesirable properties, but representing self-gravitating objects with a compactness 
as high to produce LR similar to those of BH. As mentioned above, one such possibility is a {\it regular BH}, i.e., an object endowed with an EH but without singularities in its interior.
Another possibility is an {\it ultracompact object} (UCO), which is 
a hypothetical astrophysical self-gravitating extended object more compact than a neutron star, but horizonless and globally regular capable of leading to LR.
This kind of UCOs are also referred to as BH mimickers (BHM) \cite{BHmimickers}.
Among the proposed BHM, {\it solitonic boson stars} (SBSs) are among the most extensively studied in recent years. 
SBSs are \emph{boson stars} (BSs)— localized self-gravitating solutions of the Einstein–Klein–Gordon system —
endowed with a sixth degree self-interacting potential introduced for non-topological solitons \cite{friedberg_scalar_1987}.\footnote{Other sextic potentials have been considered in  \cite{siemonsen_stability_2021, kleihaus_rotating_2005, cardoso_energy_2023}.}
This sextic potential is, however,  characterized by only two parameters: the mass term $\mu^2$, and another one, $\sigma^2$, 
responsible for controlling the strength of self-interaction.  If $|\sigma|$ is relatively low, then the self-interaction becomes high and the BS becomes an UCO. However, if that 
parameter is large, the field becomes massive but non-self-interacting, like a {\it mini boson star} (MBS).
SBSs typically attain higher compactness ${\cal C}=M/R$ (where $M$ is the 
mass and $R$ the radius of the object) than MBSs and quartic self-interacting BSs (SIBSs), often crossing the UCO threshold ($\mathcal{C}>1/3$) and thus -- unlike MBSs and SIBSs -- develop a characteristic pair of LRs, 
an outer unstable and an inner stable one. This feature also contrasts with a \Sc BH that features only one 
unstable LR. The equilibrium sequences and stability of SBSs closely mirror those of neutron and mini-boson stars— stable along the fundamental branch up to the first turning point and unstable beyond \cite{liebling_dynamical_2023}. However, SBSs may possess 
a ``second stable'' branch after a first unstable branch. It is in this second stable branch (e.g. near the global maximum mass) 
that LRs start appearing. It is due to the presence of a stable LR that some arguments have been 
put forward to claim potential gravitational instabilities in SBSs \cite{Keir2016,Cardoso2014,Benomio2021}. Numerical evidence towards this conjecture was 
first presented in the rotating case \cite{Cunha2017}, therefore questioning SBSs as potentially physically viable BHMs. Nevertheless, a subsequent non-linear perturbative analysis in spherical symmetry --supported by a linear-perturbation exploration-- showed that SBSs can be stable \cite{marks_long-term_2025}, and also a more recent study concludes that rotating SBSs may indeed be stable \cite{efstafyeva_stability_2025}.

Recent studies have mapped SBS families over wide ranges of the parameter $\sigma$ \cite{collodel_solitonic_2022}, but with $|\sigma| \gtrsim 0.039$.\footnote{\citet{collodel_solitonic_2022} employ a different convention. As a result, their solitonic coupling parameter $\sigma_{\rm CD}$ is related to ours by $\sigma_{\rm CD}=\sqrt{2\pi}\sigma$. In particular, their smallest value $\sigma_{\rm CD}=0.1$ corresponds to $\sigma\simeq 0.039$.   }In flat spacetime the sextic-potential structure supports \emph{$Q$-balls}, finite-energy lumps stabilized by a conserved Noether charge—which, when coupled to GR, become gravitationally bound BSs \cite{lee_nontopological_1992,boskovic_soliton_2022,kleihaus_rotating_2005}.

In this paper we explore SBS solutions with a parameter as low as $|\sigma| = 0.02$, which is lower than those analyzed so far (we focus only on $\sigma>0$). This is rather challenging as one requires a numerical code with more than $380$ decimals, a precision that is higher than, for instance, a quadruple-precision FORTRAN code, and thus, difficult to implement using standard number-crunching codes. Here, however, we use a code based on \textsc{Julia} with arbitrary-precision arithmetic limited only by the memory allocation (see Appendix \ref{secc:appendix_Shooting_method}).
Such low values for $\sigma$ lead to SBS models that are even more compact than those so far analyzed ($\sigma \gtrsim 0.039$), and thus they can mimic more faithfully the BH properties.
Moreover, in order to understand these UCOs in a more heuristic and simple fashion, we propose a model that mimics such an UCO 
(the mimicker of a BHM) which consists of an incompressible spherically symmetric perfect fluid in hydrostatic equilibrium with a suitable large pressure at the center (hereafter referred to as IPFUCO). In such a scenario and as the central pressure increases the IPFUCO 
starts approaching the highest possible compactness ${\cal C}= 4/9$ (the Buchdahl limit) that an extended body can have in GR, all this without the need of additional complicated and exotic assumptions (like gravastars \cite{Mazur2023}). Moreover, and which is more relevant in this context, as the compactness increases, the effective potential for null geodesics associated with this ultracompact IPFUCO develops LRs, one stable 
at the interior of the object (associated with a local minimum), and one unstable 
(associated with a local maximum) which is at the \Sc value $r=3M$, but outside the IPFUCO. 
These IPFUCOs may serve as a toy model to understand the general features 
of other UCOs like the SBS, and also as a role model for other UCOs based on field theory 
intended to reach a compactness as high as 4/9. Finally, we also analyze IPFUCOs
based on an $f(R)$ metric theory of gravity similar to the Starobinsky proposal 
for inflation, $f(R)= R + a R^2$ (hereafter $R^2$--gravity).
This is motivated because for $a \gg 1 {\rm km}^2$ one can expect objects with a maximum mass larger than those in GR 
(see Ref.~ \cite{Hodek2025} for an updated analysis), although some of these might be unstable \cite{Pretel2020}. 
In turn, this opens the possibility that an UCO based on $R^2$--gravity can be more 
compact than its counterpart in GR, and therefore that the standard Buchdahl limit in GR ${C}=4/9$ might be overcome. In particular, we explore 
this possibility with IPFUCOs. 

The paper is organized as follows. In Secs.~\ref{sec::Non rotating Boson Stars} and \ref{sec:num} we analyze thoroughly and numerically SBS models in spherical symmetry and examine the space of parameters leading to UCOs that host LRs. We also discuss some stability issues related with 
the presence of those LRs in view of of some recent developments reported in 
the literature. Sec.~\ref{sec:incfluid} analyzes IPFUCOs as simple models to understand heuristically the behavior of the UCOSBS studied in the previous section. Finally, Sec.~~\ref{sec:fR} also analyzes IPFUCOs but in the framework of $R^2$--gravity in order to contrast the possible differences with UCOs in GR. 
Two Appendices are included at the end to discuss more technical aspects related with the numerical analysis and $Q$-balls (solutions similar to SBSs but in Minkowski spacetime).

\section{Non rotating Boson Stars}
\label{sec::Non rotating Boson Stars}
The study of BSs has been extensively addressed in the literature \cite{jetzer_boson_1992, schunck_topical_2003, liddle_structure_1992,liebling_dynamical_2023}. The earliest investigations of what are known today as  MBSs
were conducted in the spherically symmetric case by \citet{kaup_klein-gordon_1968}, followed by subsequent studies by \citet{ruffini_systems_1969}, which included only a mass term in the scalar-field potential. Around two decades 
later, it was discovered that adding a quartic self-interacting term MBS models might increase its maximum mass \cite{colpi_boson_1986}. 
The first analysis of rotating MBSs in full GR was performed by Schunck 
\cite{schunck_topical_2003}, who showed that these objects exhibit toroidal topology and can also feature ergoregions. Since that pioneer 
study many analysis of rotating MBSs with or without self-interactions have been reported in the literature (for a review see \cite{liebling_dynamical_2023,delgado2025}).

Our focus of BSs is confined only to the non-rotating case and in the framework of GR. BSs are equilibrium configurations which consist of a single massive complex-valued scalar field $\Phi$ coupled to gravity. As such, the system is described by the action\footnote{We use units where $G=c=\hbar=1$ except for Sec.~\ref{sec:fR} where we restore the constants $G$ and $c$.}
\begin{equation}
S=\int d^{4}x\,\sqrt{-g} \Bigl[\frac{R}{2\kappa}
- g^{ab}\nabla_{a}\Phi^{*}\nabla_{b}\Phi
- V(|\Phi|^{2})\Bigr],
\label{actionfunct}
\end{equation}
where $\kappa = 8\pi$ and $V(|\Phi|^{2})$  is the scalar potential that is introduced below in Eq.~ \eqref{eq::SBS_Potential}. The variation of the action functional with respect to both the metric tensor $g_{ab}$ and the scalar field yields the Einstein-Klein-Gordon (EKG) system:
\begin{align}
    G_{ab} &= \kappa\,T_{ab},\\
 \nabla_a\nabla^{a}\Phi =&\Phi\frac{dV\!\left(|\Phi|^2\right)}{d|\Phi|^2},
\end{align}
for which the energy-momentum tensor (EMT) of the complex scalar field is given by  
\begin{equation}
\label{eq::EMT_EKG}
T_{ab}
= 
    \nabla_a\Phi\nabla_b\Phi^*
  + \nabla_b\Phi\nabla_a\Phi^*
  - g_{ab}\Bigl(
        g^{cd}\nabla_c\Phi\nabla_d\Phi^*
      + V\bigl(|\Phi|^2\bigr)
    \Bigr).
\end{equation}
All the conventions follow \cite{bezares_gravitational_2022, macedo_astrophysical_2013}.
Since our analysis is restricted to a static and spherically symmetric spacetime, we adopt the following metric in 
the standard (areal) spherical coordinates:
\begin{equation}
\label{eq:Lineelement}
ds^2 = -e^{\nu(r)} dt^2 + e^{u(r)} dr^2 + r^2\!\left(d\theta^2 + \sin^2\theta\, d\varphi^2\right),
\end{equation}
where $\nu(r)$ and $u(r)$ depend on radial coordinate $r$. Moreover, we assume the following ansatz for the scalar field
$\Phi(t,r) = \phi(r)e^{-i\omega t}$, where $\phi$ is a real-valued function and $\omega \in \mathbb{R}$ is the field frequency, so that the EMT \eqref{eq::EMT_EKG} is time-independent and satisfies the underlying symmetries of the spacetime.  As a result, by inserting the metric \eqref{eq:Lineelement} into the field equations, the EKG equations reduce to the following system of ordinary differential equations: 
\begin{subequations}
\label{eq::EKG_Full_system}
\begin{align}
-\frac{1}{r^{2}} + e^{-u}\left(\frac{1}{r^{2}} - \frac{u'}{r}\right) &= -\kappa \rho, \label{eq:Ein_tt}\\
-\frac{1}{r^{2}} + e^{-u}\left(\frac{1}{r^{2}} + \frac{\nu'}{r}\right) &= \kappa P_\text{rad}, \label{eq:Ein_rr}\\
\frac{e^{-u}}{2}\left(\nu'' + \frac{\nu'^{2}}{2} - \frac{\nu' u'}{2} + \frac{\nu' - u'}{r}\right) &= \kappa P_\text{tan}, \label{eq:Ein_thth}\\
\label{eq:KG}
\phi'' + \left(\frac{2}{r} + \frac{\nu' - u'}{2}\right)\phi' + e^{u}\left(e^{-\nu}\omega^{2} - \frac{d V(\phi^{2})}{d\phi^2}\right)\phi &= 0,
\end{align}
\end{subequations}
where the density, radial and tangential pressure,  $\rho$,  $P_\text{rad}$, $P_\text{tan}$, are defined through the EMT of the scalar field as
\begin{subequations}
\begin{align}
\label{densphi}
\rho &= e^{-\nu}\,\omega^{2}\,\phi^{2} + e^{-u}\,\phi'^{2} + V(\phi^{2}),\\
\label{presrphi}
P_\text{rad}&= e^{-\nu}\,\omega^{2}\,\phi^{2} + e^{-u}\,\phi'^{2} - V(\phi^{2}),\\
\label{prestphi}
P_\text{tan}&= e^{-\nu}\,\omega^{2}\,\phi^{2} - e^{-u}\,\phi'^{2} - V(\phi^{2}).
\end{align}  
\end{subequations}

Given that $P_\text{rad}$ and $P_\text{tan}$ are different in general, the BS models naturally possess an anisotropic EMT.
Note that Eq.~~\eqref{eq:Ein_thth} is not independent of the rest. In practice we use such equation as a numerical check for a consistent solution of the whole system. As stressed before, the first and simplest spherical BS models that were analyzed correspond to MBSs, which are associated with a massive but otherwise free scalar field. 
 A sequence of MBS solutions that resemble neutron stars are obtained by variations of the 
central value of the field $\phi_0$ (i.e. at $r=0$) \cite{kaup_klein-gordon_1968}. These sequences show that the total gravitational mass of the BS increases monotonically with $\phi_0$ until a maximum mass is reached, and then decreases, leading to unstable configurations. However, by introducing an additional quartic self-interacting term $\Lambda (\Psi^*\Psi)^2/4$ \cite{colpi_boson_1986}, one can construct families of maximum mass models with increasing values of $\Lambda$. This kind of MBS, while compact, does not exhibit LRs \cite{guzman_spherical_2009}. Due to this feature, and since MBSs have been extensively studied in the literature, we omit that analysis here and instead focus on SBSs, which are BSs endowed with a sixth-order potential \cite{friedberg_scalar_1987,lee_nontopological_1992}:

\begin{equation}
\label{eq::SBS_Potential}
    V\bigl(|\Phi|^{2}\bigr)
=\mu^{2}\,|\Phi|^{2}
\Bigl(1-2\,\frac{|\Phi|^{2}}{\sigma^{2}}\Bigr)^{2}= \mu^{2}\,\phi^{2}
\Bigl(1-2\,\frac{\phi^{2}}{\sigma^{2}}\Bigr)^{2}
=\mu^2\phi^2
- 4\,\mu^2\,\frac{\phi^4}{\sigma^2}
+ 4\,\mu^2\,\frac{\phi^6}{\sigma^4},
\end{equation}
where $\mu$ is the mass term, and $\sigma$ is a dimensionless parameter that controls the strength of the self-interaction 
and also the compactness of the SBS. For instance, if $|\sigma|\gg 1$ the resulting BS models reduce to MBS. On the other hand, the ultracompact BS are SBS with $|\sigma|\ll 1$. As we show in Sec.~ \ref{secc::Two_parameter_family}, for 
$|\sigma|\lesssim 0.06$ the SBSs start developing LRs, namely, in the second stable branch. This kind of SBS can serve as a potential black-hole mimicker. Alternative forms of this potential, such as a three-parameter sextic potential, have been considered to construct BS and $Q$-ball solutions \cite{kusmartsev_gravitational_1991,volkov_spinning_2002,kleihaus_rotating_2005,gervalle_chains_2022, tamaki_how_2011,boskovic_soliton_2022}. The 
theory at hand, as given by \eqref{actionfunct} and the potential \eqref{eq::SBS_Potential}, is sometimes 
rewritten by introducing factors of $1/2$ (cf. Refs.~ \cite{marks_long-term_2025,ge_hair_2024}). Here we stick to the conventions followed in Refs.~ \cite{palenzuela_gravitational_2017,bezares_gravitational_2022}. 
In most cases we can bring all such conventions together by a suitable rescaling of the scalar-field, the mass $\mu$ and the $\sigma$ parameter.

\subsection{Boundary conditions and rescaling}
\label{sec::Boundary conditions and reescaling}

  The coupled system Eqs.~\eqref{eq:Ein_tt}-\eqref{eq:KG} is actually an eigenvalue problem for the scalar field $\phi (r)$ like 
  in the MBS scenario:  given a central value $\phi_0\equiv \phi(0)$ one looks for the value $\omega(\phi_0)$ (the eigenfrequency) 
  that leads to a vanishing $\phi(r)$ as $r\rightarrow \infty$. The eigenfrequency is found from the numerical 
  solution of the system by implementing a {\it shooting method.}
  As in stationary quantum mechanical problems allowing for bound states, 
  the fundamental state (ground state) corresponds to a solution with no nodes. Here we focus only on this kind of solutions since 
  those having nodes correspond to excited states, which are unstable and decay or migrate to the fundamental solution. In this way, the globally regular (localized) SBS family is parameterized by the central value $\phi_0$.
  In order to solve the full EKG system \eqref{eq:Ein_tt}-\eqref{eq:KG} we impose suitable boundary conditions. First, we impose the following behavior near the origin:
\begin{equation}
\phi(r)=\phi_0+\frac{1}{2}\phi_2\,r^2+\mathcal{O}(r^4),\qquad
\nu(r)=\nu_0+\frac{1}{2}\nu_2\,r^2+\mathcal{O}(r^4),\qquad
u(r)=\frac{1}{2}u_2\,r^2+\mathcal{O}(r^4),
\end{equation}
so that $\phi'(0)=0$, $\nu'(0)=0$ and $u(0)=0$. Substituting into the Klein--Gordon
equation \eqref{eq:KG} we obtain at leading order,
\begin{equation}
\label{eq:origin_coeffs}
\phi_2=-\frac{1}{3}\Big(e^{-\nu_0}\omega^2 - V'(\phi_0^2)\Big)\phi_0,\qquad
u_2=\frac{2}{3}\kappa\Big(e^{-\nu_0}\omega^2\phi_0^2+V(\phi_0^2)\Big),\qquad
\nu_2=\frac{2\kappa}{3}\left(2e^{-\nu_0}\omega^2\phi_0^2- V(\phi_0^2)\right).
\end{equation}
On the other hand, in order to recover the Minkowski spacetime at spatial infinity ($r\to\infty$),
we impose the following conditions asymptotically:
\begin{equation}
\phi(r)\to 0,\qquad u(r)\to 0,\qquad \nu(r)\to 0,
\end{equation}
so that $e^{u}\to 1$ and $e^{\nu}\to 1$ at spatial infinity.  

Since the system of equations depends on $\nu$ only through its derivatives and the
combination $e^{-\nu}\omega^2$, the transformation $\nu\to\nu+\delta$
corresponds to a constant time rescaling $t\to e^{-\delta/2}t$ and leaves the
system invariant provided $\omega\to e^{\delta/2}\omega$. We exploit this
freedom to enforce the standard asymptotic gauge $\nu(\infty)=0$, so that $t$
coincides with the proper time of a static observer at infinity. In practice
we integrate for an unnormalized $\tilde\nu(r)$ up to a large radius $r_{\max}$ 
where $\phi(r)\sim {\cal O}(10^{-14}) $
and then shift
\begin{equation}
\nu(r)=\tilde\nu(r)-\tilde\nu(r_{\max})
\qquad\Longleftrightarrow\qquad
e^{\nu(r)}=\frac{e^{\tilde\nu(r)}}{e^{\tilde\nu(r_{\max})}},
\end{equation}
which guarantees $\nu(r_{\max})\simeq 0$. The corresponding ``physical" frequency
(after restoring the normalization $\nu(\infty)=0$) is $\omega\to\omega\,e^{\tilde\nu(r_{\max})/2}$.

In order to solve  numerically the system \eqref{eq:Ein_tt}-\eqref{eq:KG}, it is convenient to introduce dimensionless variables for $r$ and $\omega$ so that the reduced system becomes dimensionless (see
Appendix~\ref{sec::Dimensionless}) as well as a rescaling in the scalar field $\phi$. A suitable choice will depend on the form of the scalar potential, in particular for the SBS scenario we follow \cite{macedo_astrophysical_2013, macedo_into_2013, palenzuela_gravitational_2017} and set 
\begin{equation}
\label{eq:Reescale_variables}
\tilde r= r \Lambda\mu,\qquad
\tilde \omega = \frac{\omega}{\Lambda\mu},\qquad
\tilde \phi=\frac{\sqrt{2}}{|\sigma|}\,\phi,\qquad
\end{equation}
where $\Lambda=\sqrt{8\pi}\,\sigma$. The choice for $\tilde \phi$ is motivated by the value the field can take at the degenerate-vacua of the potential \eqref{eq::SBS_Potential} at $\phi= \pm |\sigma|/\sqrt{2}$, which corresponds to $\tilde \phi= \pm 1$. Notice that both the shift $\nu\rightarrow \nu-\tilde\nu(\tilde{r}_{\max})$
and the dimensionless rescaling \eqref{eq:Reescale_variables} involve a change
in the numerical value of the frequency. Thus, for a given central amplitude
$\phi_0$ we distinguish between the eigenvalue $\tilde\omega$ returned by the
shooting procedure for the unnormalized system, and the ``physical" dimensionless frequency ${\tilde \omega}\rightarrow \tilde\omega\,e^{\tilde\nu(\tilde{r}_{\max})/2}$  associated with the gauge
choice $\nu(\infty)=0$. In the following we always report
the final, properly normalized values.


\subsection{Global quantities}
\label{sec::Global quantities}
Due to the assumed symmetries, the ADM (gravitational) mass $M$ and the Komar mass coincide. To obtain an expression for 
such quantity, we consider the mass aspect function $m(r)$ as introduced through the standard parametrization 
in area-$r$ coordinates:
\begin{equation}
\label{eq:mass_function_def}
e^{-u(r)} \equiv 1-\frac{2m(r)}{r},
\qquad\Longleftrightarrow\qquad
m(r)=\frac{r}{2}\left(1-e^{-u(r)}\right).
\end{equation}
In this way Eq.~ \eqref{eq:Ein_tt} is recasted as a differential equation for $m(r)$, 
\begin{equation}
\frac{dm(r)}{dr}= \frac{\kappa}{2} r ^2\rho
\label{eq:mass_adm_def}
\qquad\Longleftrightarrow\qquad
M= \lim_{r\to\infty} m(r)
= 4\pi \int_{0}^{r} y^{2}\rho(y)dy.
\end{equation}
 Accordingly, with the rescale choice \eqref{eq:Reescale_variables}, we have $\tilde m= m \Lambda\mu$. We emphasize that BS,  unlike fluid stars, do not possess a sharp surface since the scalar field extends up to infinity, albeit with an exponential decay.  Therefore, the radius of the star, $R_{99}$, is usually defined as the value of $r$ enclosing $99\%$ of the total mass, i.e., $m(R_{99})=0.99 M.$\footnote{Other operational definitions of the radius are possible
\cite{collodel_solitonic_2022}, leading to slightly different numerical values
for the compactness.} With this convention the compactness can be written as $\mathcal{C}=M/R_{99}$. Note that this value is independent of $\mu$ (cf. Eq.~~\eqref{eq:Reescale_variables}).

Finally, notice that the Lagrangian of the EKG system is invariant under global phase transformations $\Phi\rightarrow \Phi e^{i\beta}$, where $\beta$ is constant. This global $U(1)$ invariance implies the existence of a conserved {\it charge} (the total boson number) associated with the corresponding Noether current,
\begin{align}
    J^{a}&=ig^{ab}(\Phi^*\nabla_b\Phi -\Phi\nabla_b\Phi^{*}),& \nabla_a J^a=0&,
\end{align}
which is conserved. The total charge is obtained by integrating the time component of this current over a spacelike hypersurface $\Sigma_t$,
\begin{equation}
\label{eq:Q-def}
Q=\int_{\Sigma_t} n_a J^a\,\sqrt{\gamma}\,d^3x
=\int_{0}^{\infty}\!\!dr\int_{0}^{\pi}\!\!d\theta\int_{0}^{2\pi}\!\!d\varphi 
\big(e^{\nu/2}\big)\big(2\,\omega\,e^{-\nu}\phi^2\big)\,e^{u/2}\,r^2\sin\theta
= 8\pi\omega \!\int_{0}^{\infty}\!dr r^2\, e^{(u-\nu)/2}\phi^2\, .
\end{equation}
Here, we have made use of the fact that the unit normal vector to the constant–time hypersurfaces, $\Sigma_t$, is given by $n^a = (e^{-\nu/2},0,0,0)$, which satisfies $n_a n^a = -1$, and of the harmonic ansatz for the scalar field, which guarantees that the spatial components $J^r$, $J^{\theta}$, and $J^{\varphi}$ vanish.\footnote{The vanishing of $J^r$, $J^{\theta}$ and $J^{\varphi}$ follows from the purely harmonic time dependence of the scalar field, which removes any net spatial flux of the Noether current.} $Q$ is commonly understood as the total number of particles (upon quantization) \cite{schunck_topical_2003}, the equivalent of the total baryon number in 
neutron stars, so that $\mu Q$ provides the total boson mass (the equivalent of the total baryon mass of a neutron star). It is more convenient to compute \eqref{eq:Q-def} numerically by introducing an auxiliary variable $q(r)$ satisfying 
the differential equation, 
\begin{equation}
\label{eq:dQdr-phys}
\frac{d q(r)}{dr}
= 8\pi\,\omega\, r^{2}\,e^{\frac{u-\nu}{2}}\,\phi^{2}, \qquad q(0)=0, \qquad Q= \lim_{r\to\infty} q(r)  ,
\end{equation}
and integrate it alongside the metric and scalar-field equation. By doing so,  we identify the total conserved charge with the value of the function $q(r)$ at the outer boundary of the numerical domain, i.e., $Q = q(r_{\max})$. 
We also compute the {\it binding energy} of the boson star from the difference of the total gravitational mass and the total boson mass:  
\begin{equation}
    E_\text{B}= M - \mu Q.
\label{ebind}
\end{equation}
The sign of $E_\text{B}$ indicates whether the configuration is energetically bound ($E_\text{B}<0$) or unbound ($E_\text{B}>0$).  Recent nonlinear simulations have shown explicitly that there exist stable SBS models with $E_\text{B} > 0$, which do not disperse or decay under small perturbations \cite{marks_perturbations_2025}. This challenges the conventional wisdom that negative binding energy is required for dynamical stability \cite{guzman_evolving_2004,guzman_spherical_2009,seidel_dynamical_1990}.  


\subsection{Light rings (LR)}
\label{sec::Light rings (LR)}
As mentioned in the introduction, a stark feature of an UCO is the presence of a \textit{photon sphere} or 
LR (which for a \Sc BH solution locates at $r=3M$). Therefore, for an horizonless UCO with mass $M$ and radius $R$ that mimics a BH 
\cite{cardoso_testing_2019}, one expects to 
satisfy the condition $R<3M$ so that a LR is within the outskirts of the UCO, and thus the compactness also satisfies $1/3< {\cal C}$. 
The LR is usually associated with circular unstable orbits for photons. 
However, due to regularity at the origin of the metric potentials, LR in horizonless UCOs always come in pairs \cite{cardoso_geodesic_2009,cardoso_light_2014, grandclement_light_2017, cunha_light_2017}. Here we find only one 
pair of LR associated with unstable (stable) circular orbits, i.e., a maximum (minimum) of the effective potential 
for null geodesics. To see this, from \eqref{eq:Lineelement} we obtain the effective potential of geodesics, $V_{\rm eff}(r)$, in the 
usual way from the conserved quantities $E = e^{\nu}\dot t$ and $L = r^{2}\dot\varphi $:
\begin{align}
\label{eq:radial_null}
e^{\nu+u}\,\dot r^{2}
&= E^{2}-V_{\rm eff}(r), & V_{\rm eff}(r)&
\equiv
e^{\nu(r)}
\left(
\varepsilon + \frac{L^{2}}{r^{2}}
\right),
\end{align}
where the parameter $\varepsilon=1,0$, for time-like and null geodesics, respectively. Here we are only interested in the 
null-geodesic problem $\varepsilon=0$. In particular,  $\dot{r}=0$ and $\ddot{r}=0$ \cite{macedo_astrophysical_2013,cardoso_geodesic_2009, Chandrasekhar} provides the conditions for circular orbits at radius  $r=r_{\gamma}$, which correspond to the location of a critical point 
of the effective potential:
\begin{equation}
\label{eq:LR_cond_general_nu_u}
\left.\frac{d}{dr}\left(\frac{e^{\nu(r)}}{r^{2}}\right)\right|_{r=r_{\gamma}}
= 0.
\end{equation}
As we show below, the effective potential for sufficiently compact SBS $(|\sigma|\ll 1)$ possesses typically a maximum (unstable LR) and a minimum 
(stable LR). While not all SBS solutions possess LR in the second stable branch (for instance models with $(0.06 \lesssim |\sigma|$), in \cite{Cunha2017} a theorem 
has been proved providing the conditions allowing for both types of LR. 
It has been argued that the mere presence of a stable LR is an indication of the unstable nature of the UCO \cite{Keir2016,Cardoso2014}, 
and perhaps more importantly, numerical evidence has been provided that this is indeed the case for some rotating SBSs 
\cite{cunha_exotic_2023}. 
Notwithstanding, it is unclear if this numerical evidence is sufficiently strong, as more recently, this conclusion has been 
questioned by several authors. In Ref.~ \cite{marks_long-term_2025} a detailed numerical analysis shows that there exist spherically symmetric 
SBS, like some analyzed here with $|\sigma|= 0.05$ (cf. Fig.~\ref{fig:sig05})\footnote{In that reference they use different variables. In order to match ours $\varphi\rightarrow \sqrt{2}\Phi$ and 
$\sigma\rightarrow \sqrt{2}\sigma$. Thus the values $\sigma=\{0.06,0.08\}$ used there correspond approximately to our values
$\sigma= \{0.042,0.057$\}, respectively. The stability analysis performed in that reference was for the 
models close to the maximal mass that hosts LRs.}, that are stable while hosting the two kind of LRs (stable and unstable) . On the other hand, in Ref.~ \cite{efstafyeva_stability_2025}, the authors show that there exist also rotating SBS 
of this kind that are also stable using the two values $\sigma= \{0.05,0.2\}$. We can thus conclude, at least prematurely, that the conjecture about the correlation between 
the instability of UCOs with the existence of stable LRs, does not always hold, as it is unclear what is the exact value of the ratio 
$V_{\rm eff}^{\rm max}/V_{\rm eff}^{\rm min}$ above which the UCO becomes unstable, and what would be the strength of the 
corresponding perturbation leading to the instability (a marginally stable object can become unstable if it is submitted 
to a sufficiently strong perturbation).

\section{Numerical results}
\label{sec:num}
%
In this section we report the numerical results of the integration of the EKG system \eqref{eq:Ein_tt}-\eqref{eq:KG} together with 
Eq.~\eqref{eq:dQdr-phys} for $q(r)$. This system of equations under the scalar-field potential \eqref{eq::SBS_Potential}
is integrated with a 4th-5th order Runge-Kutta algorithm  (a more detailed discussion can be found in Appendix~\ref{secc:appendix_Shooting_method}). The potential \eqref{eq::SBS_Potential} is the simplest form can support spatially localized, time–dependent solutions in Minkowski spacetime known as $Q$-balls (stabilized by a global $U(1)$ charge; cf. Appendix \ref{sec:Qball}).\footnote{The existence of non-topological solitons (in the absence of gravity) requires two key conditions. First, the theory must possess a global $U(1)$ symmetry. This guarantees a conserved Noether charge and, in turn, constrains the potential to depend solely on the invariant combination $\Phi^*\Phi$. Second, the potential $V\!\left(|\Phi|^{2}\right)$ must include an attractive self-interaction term. In the case of the potential \eqref{eq::SBS_Potential}, such an attractive contribution is indeed present (in the fourth order term). Moreover, to ensure that the potential remains bounded from below and that the theory stabilizes at large field amplitudes, 
one must introduce at least a sixth-order term in $|\Phi|$ \cite{lee_nontopological_1992, jetzer_boson_1992, kesden_gravitational-wave_2005}.}
 Notice that it admits a degenerate-vacuum configuration at $\phi=\pm |\sigma|/\sqrt{2}$. As a result, one can construct solutions whose ``interior" is well approximated by one of these nearly constant degenerate-vacuum state, while the ``exterior" approaches the ``true" vacuum $\phi = 0$. The two regions are separated by a thin layer across which the field rapidly interpolates between $\phi\approx |\sigma|/\sqrt{2}$ (when taking non-negative values for the field $\phi$) and $\phi\approx 0$ while crossing the potential barrier with a maximum at $\phi= \sigma/\sqrt{6}$. The parameter $\sigma$ therefore controls the characteristic interior-field amplitude, whereas the thickness of the interface is set by the inverse boson mass, $\Delta r \sim \mu^{-1}$. Due to this sharp transition, which renders Eqs. \eqref{eq:KG} stiff, the numerical integration becomes challenging as $|\sigma|\ll 1$ requiring  a high-precision numerical scheme 
in order to find the eigenfrequency $\omega$ leading to the scalar-field solution with the correct asymptotic behavior. It is for such low values of $|\sigma|$ that the mass and the compactness of the SBS can increase (relative to those of a mini BS)
when $\phi_0$ approaches the degenerate-vacuum value $|\sigma|/\sqrt{2}$. All such 
values for $\phi_0$ are associated with the second stable branch of SBS solutions leading to the most compact configurations. 
Since the minimum and maximum mass and compactness associated with that branch are found for values 
$\phi_0\sim |\sigma|/\sqrt{2}$, the mass and compactness become very sensitive to $\phi_0$ and increases abruptly as 
$\phi_0$ increases near but beyond $|\sigma|/\sqrt{2}$. Like in the first branch, the mass decreases again marking the 
beginning of the second unstable branch. It is 
in the second stable branch of SBS, where the corresponding eigenfrequencies associated with $\phi_0\sim |\sigma|/\sqrt{2}$
are extremely close to each other that higher numerical precision is required to identify each of them 
(up to 380 decimals may be required).
As stressed above, the field configuration $\phi(r)$ of this branch interpolates between $\phi_0\sim |\sigma|/\sqrt{2}$
and $\phi(r\rightarrow \infty)\rightarrow 0$ rather sharply, a fact that motivated the \textit{thin-wall} approximation where the scalar field amplitude is seen broadly as a step function \cite{kesden_gravitational-wave_2005, boskovic_soliton_2022, friedberg_scalar_1987, cardoso_eco-spotting_2022}. Within this approximation 
the scalar-field $\phi(r)$ is almost constant within the SBS around the origin $r=0$, where the field takes a value near the degenerate vacuum of the potential $\phi=|\sigma|/\sqrt{2}$, and then only within a thin-shell $\mu\Delta r \ll 1 $ located near $R_{99}$ the non-zero gradients of the field start developing, which is the region that interpolates between the degenerate-vacuum and the true vacuum of the field which is reached asymptotically ($r\rightarrow \infty$)
while crossing the potential barrier.

A commonly adopted value in studies of SBS is $\sigma=0.05$, which satisfies the condition $\sigma\ll 1$ for an UCO. This choice has been employed in a variety of contexts. For instance, \citet{macedo_astrophysical_2013, macedo_into_2013} analyzed the circular geodesic motion and its possible astrophysical signatures, while \citet{bezares_final_2017,palenzuela_gravitational_2017,bezares_gravitational_2018,bezares_gravitational_2022} and \citet{cardoso_echoes_2016} adopted the same value to construct superposed spherically symmetric initial data for binary BS simulations. Spinning configurations have been considered in \cite{cunha_exotic_2023, siemonsen_stability_2021}.  
Other choices of $\sigma$ without resorting to approximations are present in different contexts \cite{rosa_imaging_2023, rosa_polarimetry_2025, ge_hair_2024,urbano_gravitational_2019, sennett_distinguishing_2017,sukhov_pseudo-spectral_2024, russo_tidal_2025, santos_radial_2024}. To our knowledge, only recently comprehensive numerical studies across a wide range of values for $\sigma$ for SBS have been only recently reported, notably in Refs.~ \cite{collodel_solitonic_2022, boskovic_soliton_2022}. Both studies demonstrate the characteristic thin-wall behavior in the high compactness regime of SBS. 
The thin-shell behavior is a well known feature that was already present in $Q$-Balls \cite{heeck_understanding_2021}.
For instance, \citet{boskovic_soliton_2022} used a mixed approach with a semianalytical approximations 
in the ultracompact regime.\footnote{\citet{boskovic_soliton_2022} also constructed numerical solutions and found a maximum compactness without approximations in the equations for $\Lambda=\{0.186, 0.19, 0.36, 0.64, 1.10, 3.30\}$, which is equivalent to $\sigma=\{0.0371047, 0.0378995, 0.0718096, 0.1276615, 0.2194183, 0.6582548\}$ in our notation. 
The choice $\Lambda= 0.186$ was also used in \cite{russo_tidal_2025}.} On the other hand, unlike our work, \citet{collodel_solitonic_2022} used a relaxation method to solve the system of ODEs and argued that the use of shooting methods leads to instabilities. Nonetheless, as we will see below, owing to the use of a high precision numerical algorithm, we are able to find solutions with values for $\sigma$ as low as $\sigma=0.02$ while still using a shooting method.



\subsection{The two-parameter $(\tilde \phi_0,\sigma)$ family of SBS solutions}
\label{secc::Two_parameter_family}
We have constructed numerical solutions in the parameter plane $(\tilde \phi_0,\sigma)$
by increasing $\sigma$ from $\sigma=0.02$ to $\sigma=0.3$ with equally separated intervals of 
$\Delta\sigma= 0.01$ , where $\tilde \phi_0=\phi_0/( \sigma /\sqrt{2})$ is the rescaled scalar field at 
$r=0$. For a given fixed $\sigma$, we then vary 
  $\tilde \phi_0$ so that the two stable and unstable branches of SBS are clearly identified by local extrema in the $M(\tilde\phi_0)$ curve. In particular, we are mainly interested in the 
second stable branch where the ultracompact SBSs are found. We present first a sample of solutions for the value $\sigma=0.05$ 
that has been adopted in several studies and for different values of $\tilde \phi_0$ \cite{macedo_astrophysical_2013, macedo_into_2013,bezares_final_2017,palenzuela_gravitational_2017,bezares_gravitational_2018,urbano_gravitational_2019,bezares_gravitational_2022,cardoso_echoes_2016,cunha_exotic_2023, siemonsen_stability_2021,collodel_solitonic_2022,sukhov_pseudo-spectral_2024} and then plot the whole family of global properties in this plane. As mentioned before, other values for $\sigma$ have been explored in similar and different contexts \cite{boskovic_soliton_2022,collodel_solitonic_2022,marks_long-term_2025,rosa_polarimetry_2025, ge_hair_2024,urbano_gravitational_2019, sennett_distinguishing_2017,sukhov_pseudo-spectral_2024, russo_tidal_2025, santos_radial_2024,marks_boson_2025}, but as far as we are aware, this has not been done in a systematic way as we do here, and neither considering values for $\sigma$ as low as $|\sigma|=0.02$, which are the most challenging to find.
The results for $\sigma=0.05$ were calibrated using the solutions reported in \cite{bezares_gravitational_2022}.

\begin{figure}
    \centering
   \includegraphics[width=\linewidth]{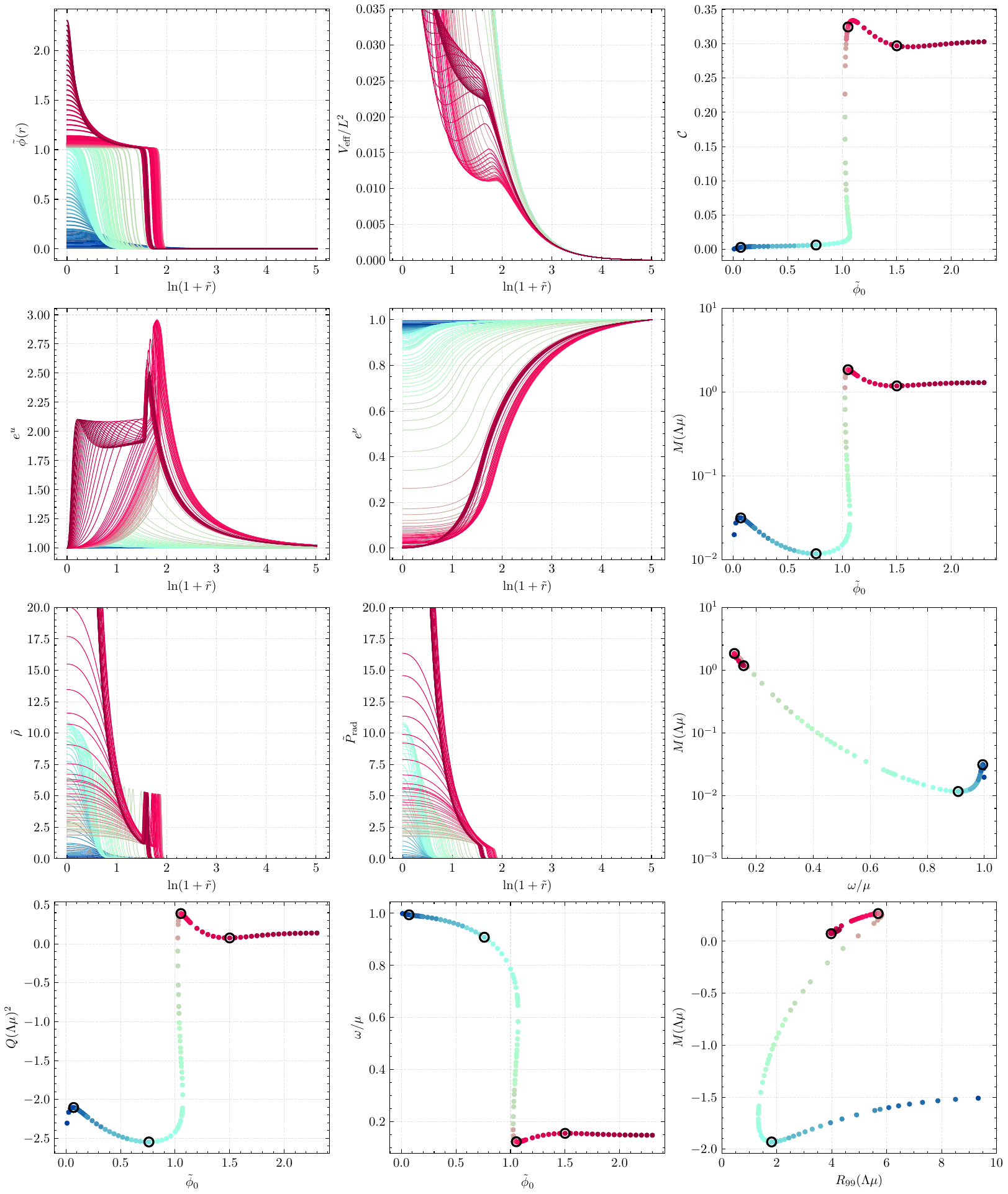}
   \caption{Radial structure and global properties for a representative sequence of SBS with $\sigma = 0.05$ 
   and different  $\tilde\phi_0$ (indicated by colors;
   rows are labeled with numbers and columns with letters). 
   Panel (1a): scalar field solution for different $\tilde\phi_0$; 
   Panel (1b): effective potential $V_{\text{eff}}/L^2=e^{\nu(r)}/r^2$. Panel (1c): compactness $\mathcal{C}$ as a function of the central field amplitude $\tilde\phi_0$. 
   Panels (2a) and (2b): metric potentials $e^{u(r)}$ and $e^{\nu(r)}$, respectively. Panel (2c): total mass $M(\Lambda\mu)$ as a function of $\tilde\phi_0$. 
   Panels: (3a) and (3b): (rescaled) energy density $\tilde\rho(r)$ and radial pressure $\tilde P_\text{rad}(r)$. 
   Panel (3c): total mass $M(\Lambda\mu)$ as a function of the field frequency $\omega/\mu$. The bottom row displays additional global quantities along the sequence: Panel (4a): Noether charge $Q(\Lambda\mu)^2$ versus $\tilde\phi_0$. Panel (4b): frequency $\omega/\mu$ as a function of $\tilde\phi_0$. Panel (4c): total mass $M(\Lambda\mu)$ as a function of the effective radius $R_{99}(\Lambda\mu)$. Open black circles mark the turning points of the mass curves $M(\tilde\phi_0)$--Panel (2c)--, which are replicated across the corresponding panels.} 
    \label{fig:sig05}
\end{figure}
Figure \ref{fig:sig05} (Panel 1a) shows the behavior of the scalar field $\tilde \phi (r)$ for several central amplitudes
$\tilde \phi_0$ (cf. bottom panel of Fig.~~1 in Ref.~ \cite{bezares_gravitational_2022}). When $\tilde \phi_0$ is relatively low compared with $\tilde \phi=1$ 
--the location of the degenerate-vacuum-- (blue to green lines), the scalar field solutions lie in the {\it thick shell} regime, where the gradients of the field are mild and spread over a broad radial domain. This behavior is correlated with both a low mass and compactness and  high field frequency  (blue–to–green markers in the third–column panels: $\mathcal{C}(\tilde{\phi}_0)$, $M(\tilde{\phi}_0)$, and $M(\omega/\mu)$;  cf. top panel of Fig.~1 in Ref.~\cite{bezares_gravitational_2022}). In fact, along this branch—up to the second low–amplitude open black circles in the $M(\tilde\phi_0)$ diagram (2b)—SBSs effectively behave as \emph{mini} boson stars. The field remains dilute ($\tilde\phi\ll 1$ everywhere), so self–interaction effects are negligible and the scalar potential can be linearized, $V(\phi^{2})\simeq \mu^{2}\phi^{2}$. Consequently, the KG equation is essentially linear with $\omega/\mu \to 1$, indicating large decay length $k^{-1}=(\mu^{2}-\omega^{2})^{-1/2}$. This can also be observed from the central energy-density (pressure) from panels 3a (3b) which is relatively small  (blue-green colors).

As $\tilde \phi_0$ increases and reaches values around $\tilde \phi=1$ (magenta and purple colors of Panel 1a), the field develops a {\it thin shell} 
(cf. Refs.~\cite{boskovic_soliton_2022,collodel_solitonic_2022}) and behaves like a step function, where the field remains almost constant and then abruptly decreases within a thin shell --where the largest gradients are located-- before reaching the asymptotic value associated with the true vacuum $\tilde \phi=0$. The solutions with a thin-shell produce a sharp transition to the second stable branch of SBS where the total mass and 
compactness increase again and abruptly with $\tilde \phi_0$ (magenta and purple dots of Panels 1c, 2c and 3c; cf. top panel of Fig.~ 1 in Ref.~\cite{bezares_gravitational_2022}). As already noted in previous works \cite{macedo_astrophysical_2013, bezares_final_2017, collodel_solitonic_2022}, it is precisely in this domain of $\tilde \phi_0$ that  determining the eigenvalues $\omega$ leading to the corresponding asymptotic behavior of each solution becomes numerically delicate and requires a higher numerical precision in the shooting method (a precision of around 34 decimal places).\footnote{For $\sigma=0.02$ we obtained convergent solutions using up to 1280-bit arithmetic; see Appendix \ref{secc:appendix_Shooting_method}. } Once the (global) maximum mass of the second branch is reached, another second unstable branch develops as one can appreciate from the $M(\tilde{\phi}_0)$ and $\mathcal{C}(\tilde{\phi}_0)$ curves (Panels 1c, 2c). After this local maxima the lowest frequency is reached and the characteristic ``spiral" behavior of BS starts.

Panels 3c and 4c plot the mass versus the radius $R_{99}$ and the frequency $\omega$, respectively, where the different branches (stable and unstable) are more easily contrasted. Finally, Panel 2a depicts the effective potential \eqref{eq:radial_null} per squared-angular-momentum $V_{\text{eff}}/L^2$ associated with null geodesics ($\varepsilon=0$). We appreciate that for low 
compactness (green curves; the corresponding blue curves 
of Panel 1a are not displayed) the potential does not develop 
extrema. As $\tilde \phi_0$ increases the scalar-field develops 
a thin shell and the effective potential shows
a minimum (stable circular orbits) and maximum (unstable circular orbits). These extrema are associated with LR. In particular, the unstable ones are those that are identified with the LR in actual astronomical observations. For instance, photons emitted from an accretion disk around the compact object can turn several times before being swallowed or scattered towards the observer forming the peculiar pattern reported by the EHT \cite{EHT1,EHT2}. 
The effective potentials show the infinite barrier at $r=0$ which is a feature typical of extended (globally regular) objects, indicating that photons with non-zero $L$ never reach the center of the SBS, but only approach it before being scattered away towards infinite. Such feature contrasts with the \Sc BH scenario, 
where the corresponding potential has only a maximum at $r=3M$ and vanishes at the horizon at $r=2M$. In that case 
the ingoing photons with energy $E^2> V_{\text{eff},{\rm max}}^{\rm Sch}$ can easily cross the event horizon until the singularity is reached 
but they never come out. Finally, for even larger 
$\tilde \phi_0$ the second unstable branch is reached (beyond the global maximum mass model--indicated with a black circle in the purple region in Panel 2c-- where the mass decreases) and the extrema of the effective potential disappear (absence of LRs). Such potentials are associated with the corresponding purple curves of Panel 1a where the scalar-field ``loses" the thin-shell 
($\tilde \phi_0 \gtrsim 1.2$).

\begin{figure}
        \centering
        \includegraphics[width=1\linewidth]{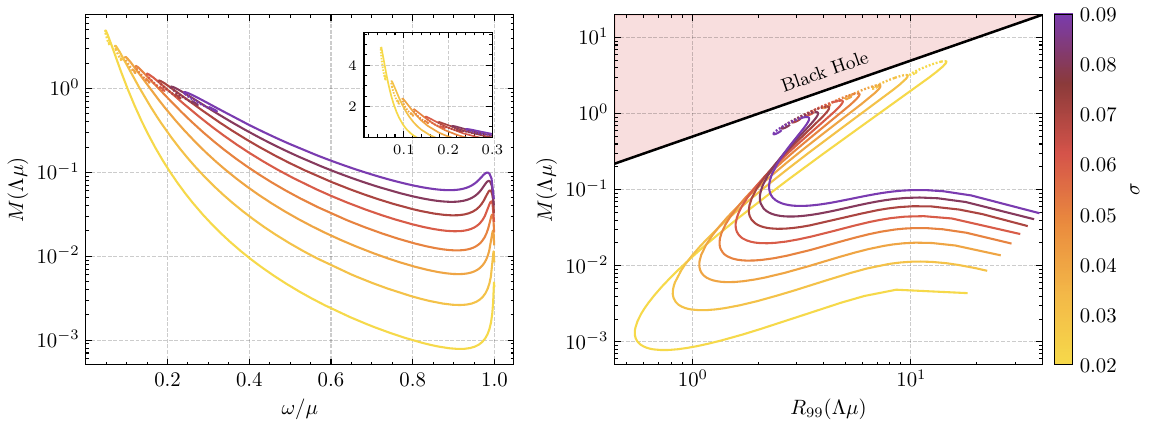}
      \caption{(Left) Domain of existence of SBSs for different values of the solitonic parameter $\sigma$. (Right) Corresponding mass–radius relations $M$–$R_{99}$, together with the diagonal line marking the black–hole limit; the shaded region above this line is excluded by $\mathcal{C}>1/2$. The colour bar encodes the value of $\sigma$. Dotted lines indicate configurations with LRs.}
\label{fig:MR_CvsR_sigma}
    \end{figure}

\begin{figure}
\centering
    \includegraphics[width=1\linewidth]{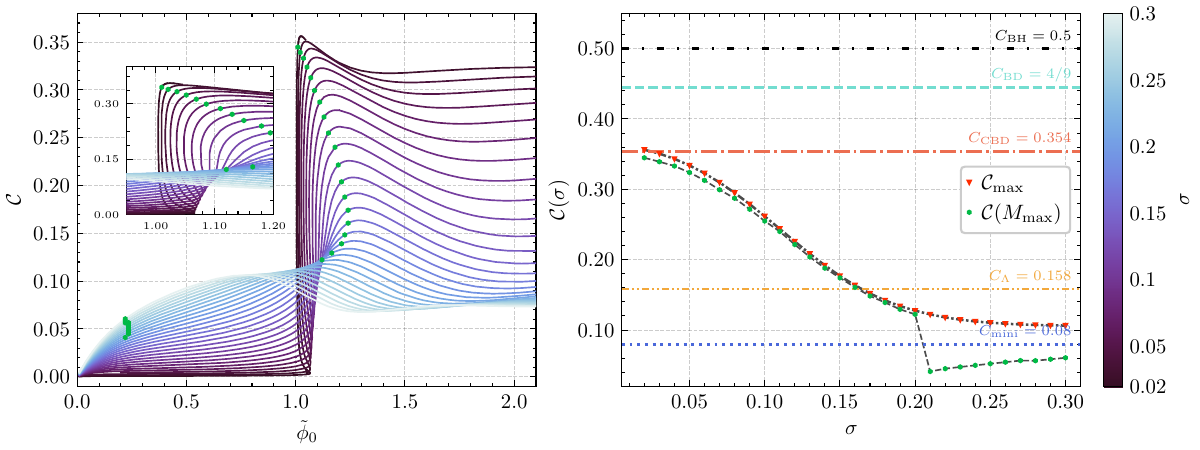}
\caption{
Compactness diagnostics for SBS sequences parameterized by the parameter $\sigma$.
\textit{Left panel:} compactness $\mathcal{C}$ as a function of the (rescaled) central scalar amplitude $\tilde\phi_0$ for each value of $\sigma$ (darker colors indicate smaller $\sigma$). The hexagon green markers indicate the configuration of maximum mass $M_{\max}$ along each sequence.
An inset highlights the region around $\tilde\phi_0\simeq 1$ where several sequences backbend indicating that $\tilde\phi_0$ no longer uniquely determine the solutions.
\textit{Right panel:} Relevant compactness solutions as a function of $\sigma$: inverted triangles denote the maximum compactness $\mathcal{C}_{\max}(\sigma)$ attained along each sequence, while hexagons denote the compactness evaluated at the maximum-mass configuration, $\mathcal{C}(M_{\max})$.
Horizontal lines indicate compactness reference values (Buchdahl bound $\mathcal{C}_{\rm BD}=4/9$, causal Buchdahl-bound $\mathcal{C}_{\rm CBD}= 0.354$, black-hole limit 
$\mathcal{C}_{\rm BH}= 1/2$, the $\Lambda$-threshold $\mathcal{C}_{\rm \Lambda}= 0.158$  \cite{amaro-seoane_constraining_2010}, and the mini-boson-star reference $\mathcal{C}_{\rm mini}= 0.08$ \cite{kaup_klein-gordon_1968}).
}
\label{fig:CompactnessLeft_MaximumcompactnessRight}
\end{figure}

Solutions with $|\sigma|< 0.05$ have a similar qualitative behavior, except that those lying in the second branch (notably the second stable branch) become even more difficult to find and require an even higher precision that those for 
$|\sigma|= 0.05$. Figures \ref{fig:MR_CvsR_sigma} and \ref{fig:CompactnessLeft_MaximumcompactnessRight} summarize the solutions 
for different values in the plane $(\tilde \phi_0,\sigma)$. Each line is associated with a fixed $|\sigma|$ (from lowest-- yellow-- to highest values--violet--). The left panel of Fig.~~\ref{fig:MR_CvsR_sigma}
shows that the (global) maximum mass (for a given $|\sigma|$) increases as $|\sigma|$ decreases (the self-interaction potential 
of the field increases), while the right panel shows the corresponding mass-radius curves. Figure \ref{fig:CompactnessLeft_MaximumcompactnessRight} (left panel) shows the compactness of the SBS models, which also increases abruptly as the second stable branch forms, and its value associated with the 
(global) maximum-mass model (green marks) also increases as $|\sigma|$ decreases. The inset  shows a characteristic backbend in the lower $\sigma$ values (a behavior already highlighted in earlier studies, e.g.  Refs.~\cite{boskovic_soliton_2022,bezares_gravitational_2022,russo_tidal_2025}).
The lowest value to present a backbend is $\sigma\sim 0.1$. Notice that the maximum compactness (for a given $|\sigma|$ in the second branch) corresponds to an unstable SBS (i.e. the global maximum-mass model does not correspond to the maximum compactness). The right panel of Fig.~\ref{fig:CompactnessLeft_MaximumcompactnessRight} shows a rather interesting trend for the compactness of SBS relative to other types of BS and compact objects. The red-triangles and green-hexagons indicate the maximum compactness and the compactness of (global) maximum-mass models, respectively. It is apparent that the 
compactness increases as $|\sigma|$ decreases below $|\sigma|\lesssim 0.2$. For $|\sigma|\gg 1$ one approaches the mini-boson star model (massive but free scalar field) whose maximum compactness is ${\cal C}_{\rm mini}\sim 0.08$ (blue dotted line), and clearly below the most compact SBS. For reference we include also the 
maximum compactness ${\cal C}_\Lambda \sim 0.158 $ (orange dash-dotted line) for the mini-boson-star models endowed with a self-interacting potential $\Lambda \phi^4$.\footnote{This $\Lambda$ is not to be confused with the scale introduced in \eqref{eq:Reescale_variables}.} The mini-boson stars do not produce LRs, while all the ultracompact SBS (e.g. $|\sigma|\le 0.05$) always produce LRs. This plot also reveals that 
for $|\sigma|\ll 1$ the SBS reaches the largest possible compactness ${\cal C}_{\rm CBD}\sim 0.354$ (pink dash-dotted line)
allowed for a perfect-fluid object under causality constraints (subluminal 
sound-speed propagation) and radial stability \cite{lindblom_limits_1984,urbano_gravitational_2019,boskovic_soliton_2022,alho_compactness_2022}. 
Relaxing the stability requirement allows for slightly more compact causal configurations, up to ${\cal C} \sim 0.364$.
These values for compactness are near but still below to the Buchdahl limit ${\cal C}_{\rm BD}= 4/9$ (cyan dashed line) 
\cite{buchdahl_general_1959}, which can be reached by an incompressible fluid star model with an arbitrary large pressure at its center 
(cf. Sec.~\ref{sec:incfluid}). As we illustrate in the next section, the ultracompact SBS, which are those with a scalar-field solutions resembling a step function, can be thus mimicked by an incompressible fluid star. The difference is that the latter has, by construction, an energy-density with a discontinuity at the star radius, while for the former, the energy-density, which is given in terms of the radial gradients of the field and the scalar-field potential, is 
almost constant within the SBS, but as the thin-shell is approached, the energy-density increases abruptly (due to the large gradients of the field) and then decreases and vanishes rapidly as the field reaches the true-vacuum solution $\tilde \phi=0$ asymptotically.


\begin{figure}
    \centering
    \includegraphics[width=1\linewidth]{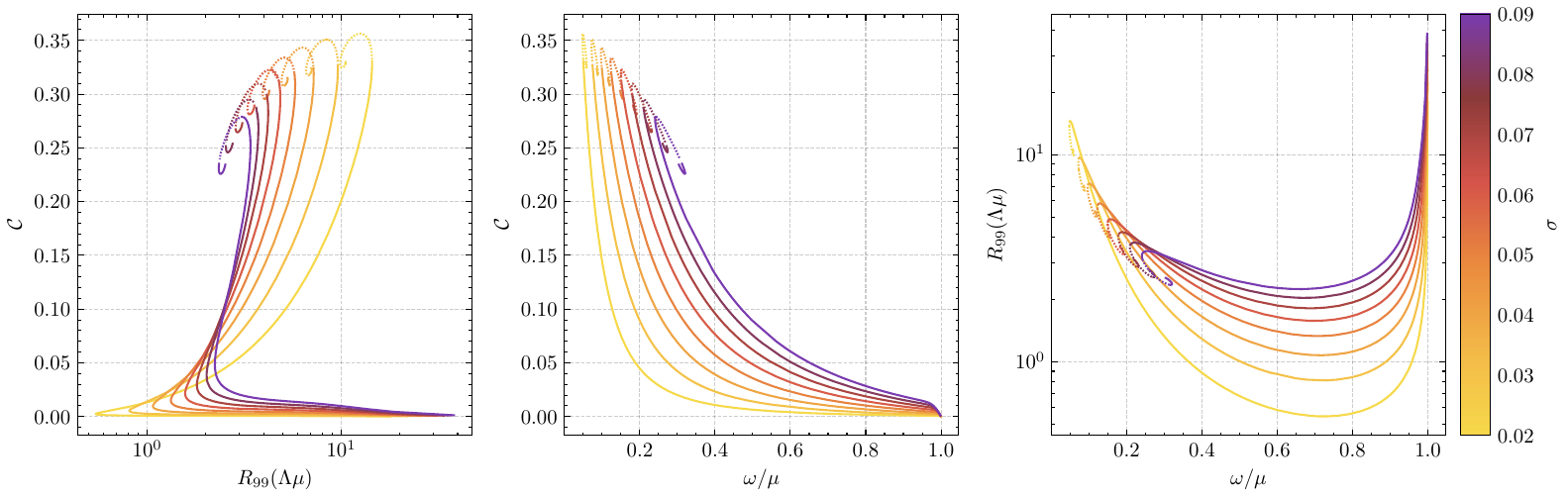}
   \caption{Compactness--radius--frequency relations for SBS. 
(Left) Compactness $\mathcal{C}$ as a function of the effective radius $R_{99}(\Lambda\mu)$. 
(Middle) $\mathcal{C}$ versus the dimensionless scalar field frequency $\omega$. 
(Right) $R_{99}(\Lambda\mu)$ versus $\omega$. 
Each curve corresponds to an equilibrium sequence with fixed coupling $\sigma$, with color encoding the value of $\sigma$ as indicated by the vertical color bar. Dotted lines indicate configurations with LRs. 
}
\label{fig:compactness-vs-several}
\end{figure}

\begin{figure}
    \centering
    \includegraphics[width=1\linewidth]{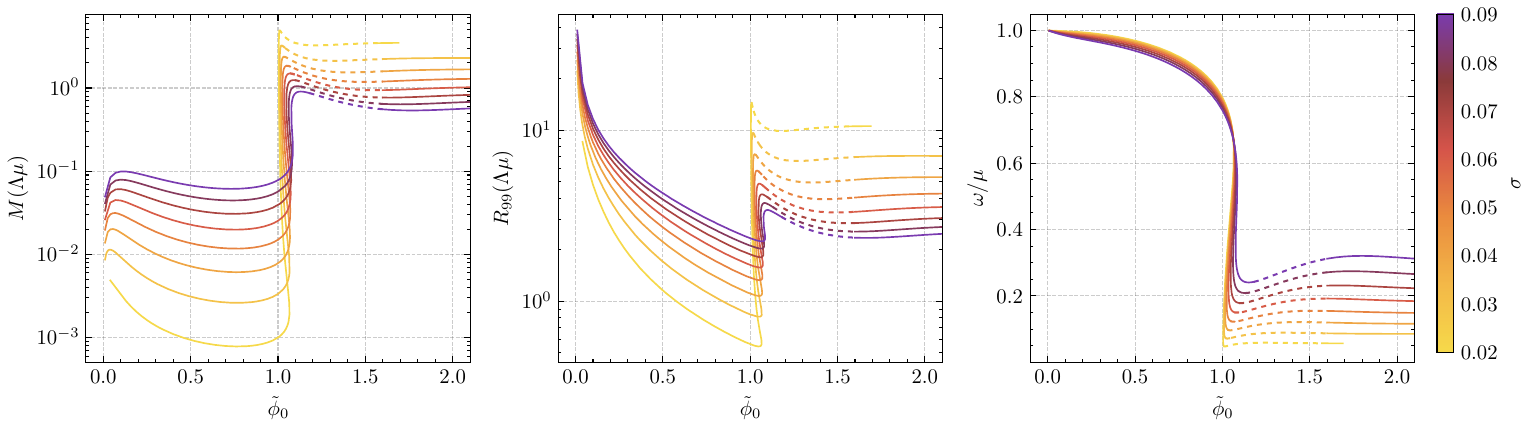}
   \caption{Mass, radius and frequency of SBS equilibrium sequences as a function of the central scalar amplitude $\tilde{\phi}_0$. (Left) Mass (rescaled by $\Lambda\mu$). (Middle) Radius $R_{99}(\Lambda\mu)$ enclosing $99\%$ of the total mass. (Right) eigenfrequency $\omega/\mu$. Each curve corresponds to a fixed value of the coupling parameter $\sigma$ (indicated by the color bar). Mass and radius are plotted on logarithmic scales, whereas $\omega/\mu$ is shown on a linear scale. Note that the three panels share a sharp change near $\tilde\phi_0=1$. Dotted lines indicate configurations with LRs.}
\label{fig:mr_soliton_sigma006_rescaled}
\end{figure}

\begin{figure}
        \centering
         \includegraphics[width=1\linewidth]{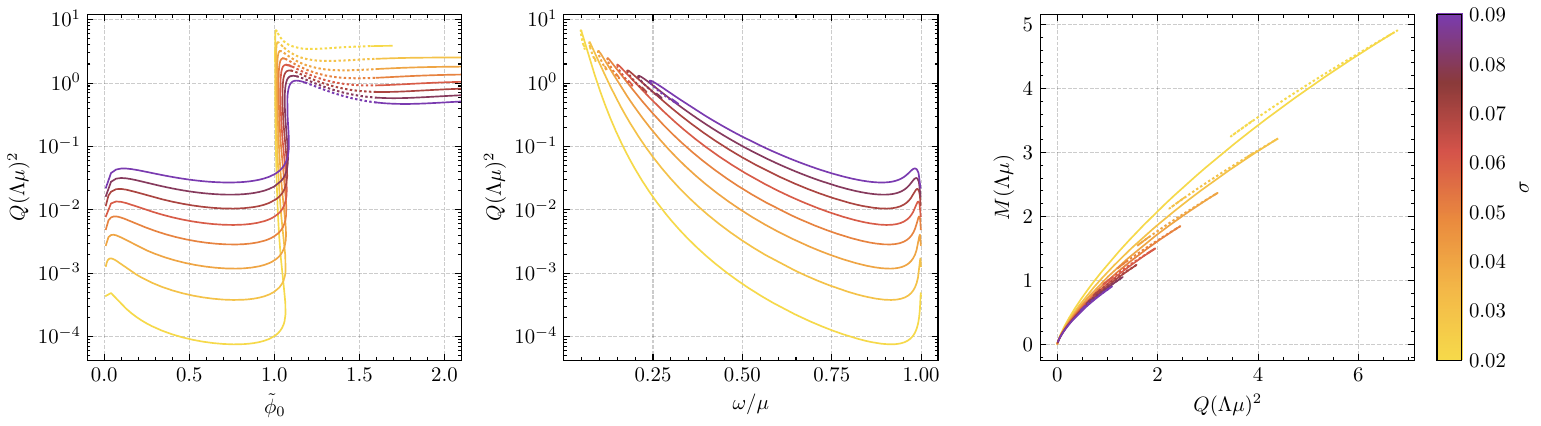}
      \caption{(Left) Noether charge  $Q(\Lambda\mu^2)$ as a function of the central scalar amplitude
$\tilde{\phi}_0$, illustrating the domain of existence of equilibrium
solutions for different values of $\sigma$. 
(Middle) Noether charge $Q(\Lambda\mu^2)$ as a function of the scalar field frequency
$\omega/\mu$, highlighting the way the total particle number
changes along each one-parameter family. 
(Right) Mass--charge relation $M(Q)$ for the same set of models, showing the behavior of the SBS where each family reaches its most massive configuration. Dotted lines indicate configurations with LRs.}
\label{fig:Q_Threepanel_charge}
    \end{figure}
\begin{figure}
    \centering
    \includegraphics[width=1\linewidth]{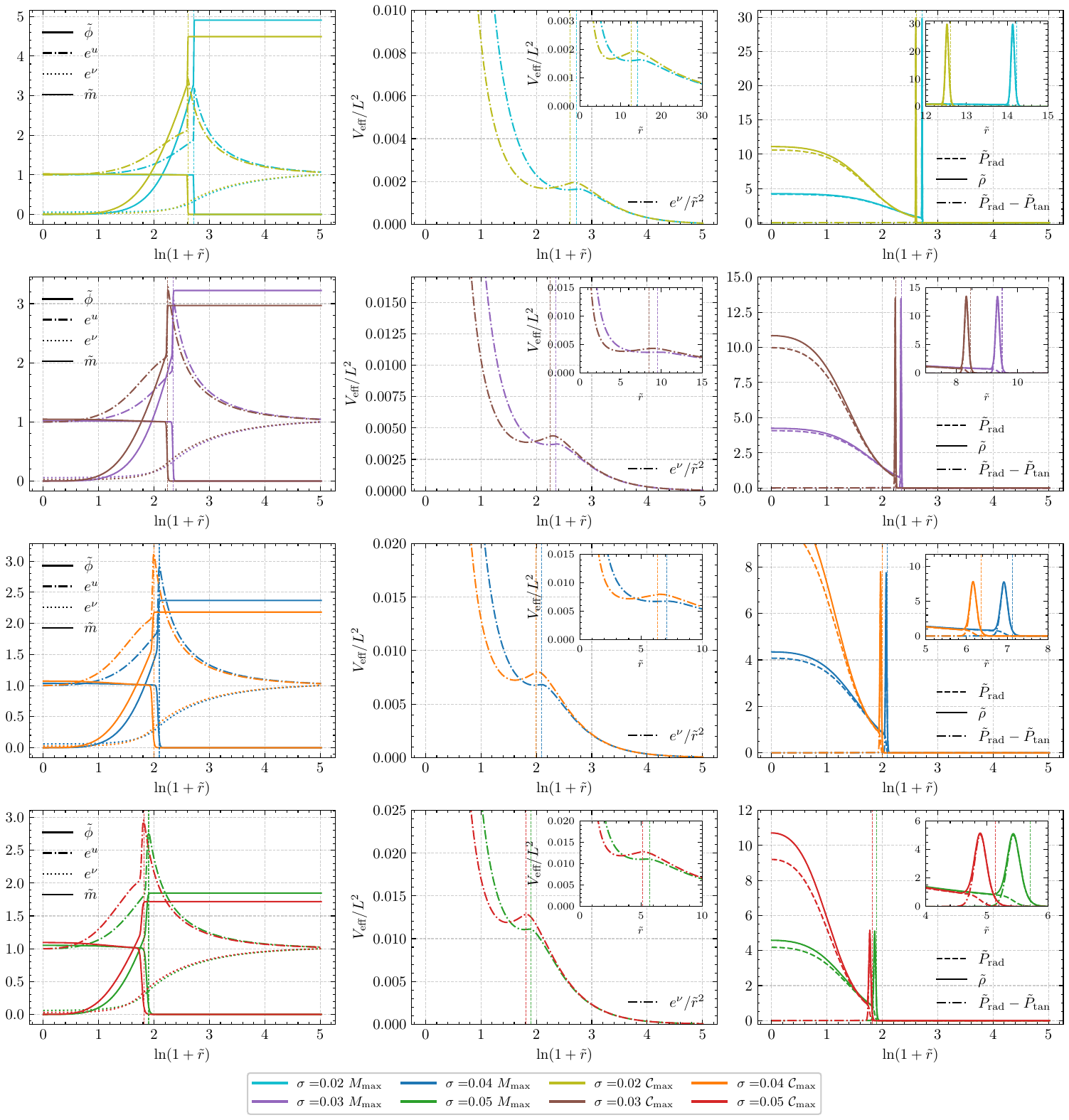}
   \caption{Radial profiles of representative SBS configurations for distinct values of the parameter $\sigma$ (from the top row with $\sigma = 0.02$ to the bottom row with $\sigma = 0.05$, cf. Table \ref{tab:extremes_sigma}). For each $\sigma$ we show the configuration with maximal mass $M_{\max}$ and the configuration with maximal compactness $\mathcal{C}_{\max}$; the corresponding color pairs are indicated in the legend at the bottom of the figure. From left to right we plot (i) the metric functions $e^{u}$ and $e^{\nu}$, (ii) the effective potential for null geodesics $V_{\rm eff}/L^{2}$ and (iii) the dimensionless pressure, density and pressure anisotropy $\tilde{P}_{\rm rad}-\tilde{P}_{\rm tan}$, all as functions of $\ln(1+\tilde r)$. Notice that for this configurations (which are compact) the pressure anisotropy effectively vanishes inside the star implying that the object behaves roughly like isotropic matter. The vertical dashed lines mark, for each curve, the radius $R_{99}$ enclosing $99\%$ of the total mass. All quantities are expressed in the dimensionless units defined in Sec.~~\ref{sec::Boundary conditions and reescaling}.}
    \label{fig::master_profiles}
\end{figure}
\begin{figure}
    \centering
    \includegraphics[width=\linewidth]{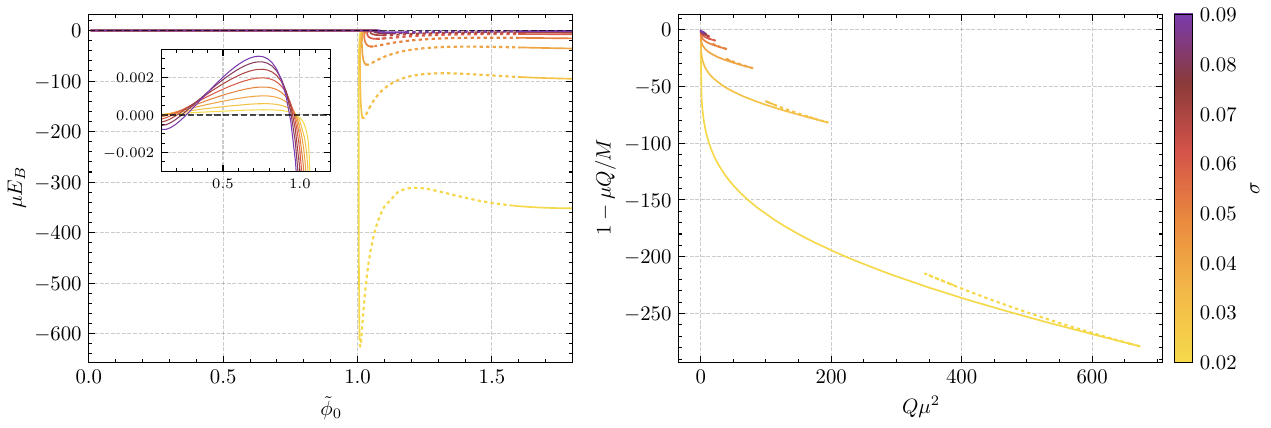}
    \caption{Binding energy and fractional binding energy for SBS sequences. \emph{Left panel:} Binding energy $\mu E_\text{B}$ as a function of the central scalar amplitude $\tilde{\phi}_0$. \emph{Right panel:} Charge–to–mass ratio, plotted as $1 - \mu Q/M$, as a function of the rescaled Noether charge $Q\mu^2$. Each curve corresponds to a fixed value of the self–interaction parameter, with the color encoding $\sigma$. The inset in the left panel zooms into the region where $\mu  E_\text{B}>0$. Dotted lines indicate configurations with LRs.  }
    \label{fig:bidnerg}
\end{figure}
\begin{figure}
    \centering
    \includegraphics[width=\linewidth]{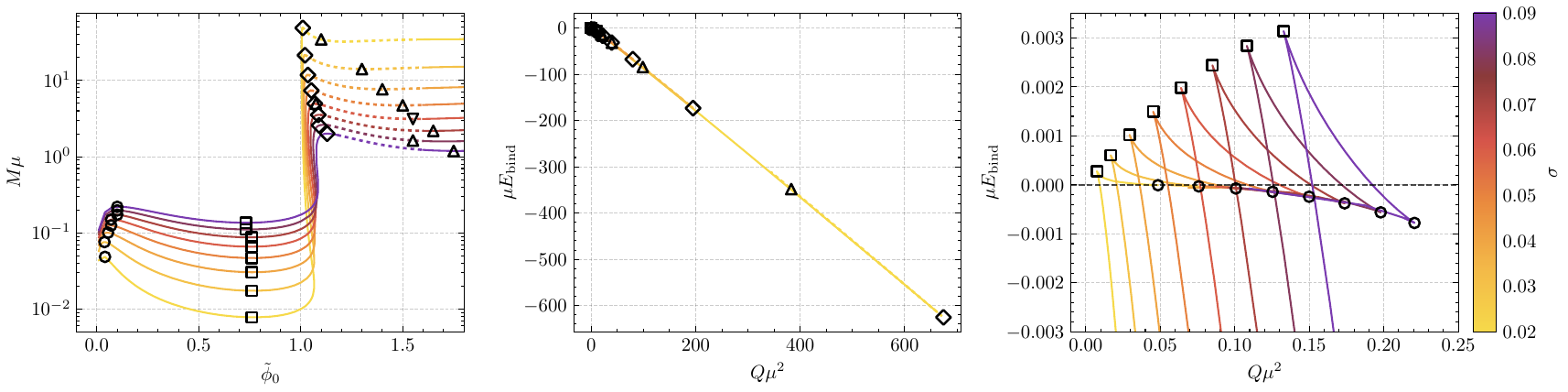}
    \caption{\textit{Bifurcation diagram for SBS.} (Left) $M(\tilde\phi_0)$ for reference. (Middle) Binding energy $\mu E_\text{B}$ versus the (rescaled) Noether charge $Q\mu^2$. (Right) $\mu E_\text{B}$ in the weakly bounded regime (zoom in of Middle panel when $Q\mu^2)\ll 1$), which corresponds to the high–frequency limit $\omega/\mu \to 1$, where a bifurcation structure emerges. The cusp markers in all three panels indicate the turning points of $\mu E_\text{B}(Q)$.}
    \label{fig:Ebind_C_Q_zoom}
\end{figure}

Figures \ref{fig:compactness-vs-several} and \ref{fig:mr_soliton_sigma006_rescaled} depict plots similar to 
Figures \ref{fig:MR_CvsR_sigma} and \ref{fig:CompactnessLeft_MaximumcompactnessRight} but with respect to other properties. 
Figure \ref{fig:Q_Threepanel_charge} shows the total Noether charge $Q$ (boson number) 
computed from \eqref{eq:dQdr-phys}, associated with values in the plane $(\tilde \phi_0,\sigma)$. 
As illustrated in the left panel, $Q$ behaves like the total mass (left panel of Fig.~~\ref{fig:mr_soliton_sigma006_rescaled}). From the two left panels of 
Figs. \ref{fig:mr_soliton_sigma006_rescaled} and \ref{fig:Q_Threepanel_charge} we clearly appreciate 
the different stable (increasing $M$ and $Q$) and unstable (decreasing $M$ and $Q$) branches, in particular, 
the sharp transition from the first unstable branch to the second stable branch of SBS around $\tilde \phi_0 \sim 1$ 
is apparent. The right panel of Fig.~~\ref{fig:Q_Threepanel_charge} depicts the total mass with respect $Q$ for different $\sigma$ where we can also distinguish the 
transition from the second stable to the second unstable branch, where the branching point indicates the value of the maximum mass and boson number of SBS. A further discussion about the stability of SBS and its relationship with the different branches appearing 
in these plots can be consulted in Refs.~\cite{ge_hair_2024,kleihaus_stable_2012,siemonsen_stability_2021}. As indicated by dotted lines in Figures \ref{fig:MR_CvsR_sigma}--\ref{fig:Ebind_C_Q_zoom}, LRs are present for 
$\sigma \gtrsim 0.08$ but this happens only in the second unstable branch of SBS.

Figure \ref{fig::master_profiles} presents explicit solutions for low $\sigma$
($ 0.02 \leq \sigma \leq 0.05$) which are associated with the configurations of 
(global) maximum mass and maximum compactness (both in the same panel for comparison). Table \ref{tab:extremes_sigma}
 displays the values of the maximum quantities. We emphasize that the 
 compactness associated with the (global) maximum mass models is below the 
 maximum compactness models (which lies in the second unstable branch slightly 
 to the right to the maximum mass model). From the three panels of first column we appreciate that $\tilde \phi_0\sim 1$ and that in all the cases the scalar field behaves close to a step function, and this behavior exacerbates as $\sigma\ll 1$. The three panels of the second column depicts the effective potential 
for null geodesics and show that the potentials for both type of configurations develop LRs, although the latter 
are more sharper for the maximum compactness configuration than those for the maximum mass. The third column 
shows the energy-density and some pressure profiles. For both kind of configurations the energy-density (the sum of 
the squares of the radial gradients of the scalar-field plus the scalar-field potential --cf. Eq.~\eqref{densphi}--) shows an approximate constant behavior for $0\leq \tilde r \lesssim 1 $ and then slowly decreases until the {\it thin shell} region is reached --where the scalar-field gradients are large-- and so then grows rapidly. Finally it falls-off also abruptly as the scalar-field vanishes outside the thin shell. The radial and tangential pressures (which are composed by similar expressions in terms of the scalar-field modulo some opposite signs in the potential $V(\phi)$ --cf. Eqs.\eqref{presrphi} and \eqref{prestphi}-- ) behave similarly.


\begin{table*}[t]
\centering
\setlength{\tabcolsep}{3pt}
\renewcommand{\arraystretch}{1.15}
\begin{tabular}{c|cccccc|cccccc}
\hline\hline
 & \multicolumn{6}{c|}{\textbf{Maximum compactness}} & \multicolumn{6}{c}{\textbf{Maximum mass}} \\
\cline{2-7}\cline{8-13}
$\sigma$ 
& $\phi_0/(\sigma/2)$ & $\omega/\mu$ & $M(\Lambda\mu)$ & $R_{99}(\Lambda\mu)$ & $Q(\Lambda\mu)^2$ & ${\cal C}$
& $\phi_0/(\sigma/2)$  & $\omega/\mu$ & $M(\Lambda\mu)$ & $R_{99}(\Lambda\mu)$ & $Q(\Lambda\mu)^2$ & ${\cal C}$ \\
\hline
0.02 & 1.024 & 0.0487482 & 4.49225 & 12.6105 & 5.88301 & 0.356231&1.01 & 0.0481763 & 4.9078 & 14.2326 & 6.77785 & 0.344828\\
0.03 & 1.046 & 0.0730346 & 2.96847 & 8.46384 & 3.86597 & 0.350724  &1.021 & 0.072292 & 3.22218 & 9.49867 & 4.40639 & 0.339224\\
0.04 & 1.072 & 0.0984092 & 2.18077 & 6.35245 & 2.80268 & 0.343296 &1.035 & 0.0972689 & 2.36775 & 7.12873 & 3.19483 & 0.332142 \\
0.05 & 1.095 & 0.124466 & 1.71751 & 5.14299 & 2.18327 & 0.333951 & 1.052 & 0.123197 & 1.84796 & 5.70489 & 2.45223 & 0.323926 \\
0.06 & 1.12 & 0.152024 & 1.39612 & 4.3252 & 1.74383 & 0.322787 & 1.069 & 0.150687 & 1.49697 & 4.78683 & 1.94754 & 0.312727 \\
0.07 & 1.146 & 0.181189 & 1.16162 & 3.75017 & 1.41965 & 0.309752 & 1.086 & 0.179869 & 1.24347 & 4.1536 & 1.58122 & 0.299372 \\
0.08 & 1.17 & 0.211761 & 0.987771 & 3.34794 & 1.18006 & 0.295038 & 1.11 & 0.209622 & 1.0536 & 3.66713 & 1.30689 & 0.287311 \\
0.09 & 1.19 & 0.243569 & 0.856891 & 3.07264 & 1.00137 & 0.278878 & 1.13 & 0.241885 & 0.905948 & 3.3342 & 1.09354 & 0.271713 \\
\hline\hline
\end{tabular}
\caption{\textit{Characteristic properties of  the maximum mass and maximum compactness configurations. }
For each $\sigma$ we report the central field amplitude $\phi_0$, the frequency $\omega$, the ADM mass $M$, the 99\% mass radius $R_{99}$, the Noether charge $Q$, and the compactness ${\cal C}$.}
\label{tab:extremes_sigma}
\end{table*}



Let us now discuss the the binding energy $E_\text{B}$ of SBS given by \eqref{ebind}. The right-panel of Fig.~\ref{fig:bidnerg} allows us to appreciate the correlation between this quantity and the stability properties 
of the SBS, notably for the second branches. For instance, for a fixed $Q$, the configurations of different $M$ belonging to each branch that minimize the binding energy are those with lower $M$, and correspond to the configurations of the (second) stable branch (cf. right panel of Fig.~~\ref{fig:Q_Threepanel_charge}). This kind of {\it bifurcation diagrams} and its relationship with stability (via catastrophe theory) had been presented in the past within the context of MBS \cite{kusmartsev_gravitational_1991,friedberg_mini-soliton_1987} 
and other SBS with generalized sextic potentials \cite{kleihaus_stable_2012}. More recently and in the current framework of SBS,  
the stability of these objects at the non-linear level has been discussed in connection with the binding energy, $E_\text{B}$,  arguing that SBS with positive $E_\text{B}$ are stable contrary to the {\it conventional wisdom} \cite{marks_perturbations_2025}. The SBS configurations with 
$E_\text{B}>0$ are those present in the sector $\tilde \phi_0 <1$ of the left-panel of Fig.~\ref{fig:bidnerg} 
and in the tiny part of the right-panel of Fig.~\ref{fig:bidnerg} where $Q \mu^2 \ll 1 $ and $  0< E_\text{B}\mu  \ll 1 $. 
The reason behind $E_\text{B}>0$ and stability can be traced back to the fact that in the low gravity regime, in particular, 
in flat spacetime, these objects exist as $Q-$balls (see Appendix \ref{sec:Qball}) where the gravitational binding energy 
vanishes. More explicitly, one expects heuristically and naively that 
$M-\mu Q= E_\text{B} <0$ due to the gravitational binding energy {\it if} $E_\text{B}= {\cal O} (G)$ where $G$ is the Newton's gravitational 
constant. In the flat spacetime background (test field) approximation
strictly speaking $M\equiv 0$ and $E_\text{B}^\text{grav} \equiv 0$ but $Q\neq 0$ (although $\mu^2 Q \ll 1$) 
since there exists the non-trivial $Q-$ball solution. In reality, one has $M= \mu Q + E_{\rm kin} + E_{\rm int }+  E_{\rm B}^\text{grav}  $, 
where $\mu Q$ is the total {\it rest} mass of bosons, $E_{\rm kin}$, the {\it kinetic} energy due to the spatial gradients of the 
(non-trivial) field configuration, $E_{\rm int }$ the {\it interaction} energy of bosons due to the scalar-field field potential 
terms $\phi^4$ and $\phi^6$, and finally the gravitational binding energy $E_\text{B}^\text{grav}= {\cal O} (G) $. The first three terms of this decomposition are expected to be positive since they come from an 
integral that involves the kinetic term plus the integral of the total scalar-field potential $V(\Phi)$, which are both 
non-negative definite. Therefore one could define $E_\text{B}= M- \mu Q= E_{\rm kin} + E_{\rm int }+  E_\text{B}^\text{grav}$. Since in the weak gravitational 
limit $|\mu E_\text{B}^\text{grav}|\ll 1 $, one can still have $E_\text{B}=  E_{\rm kin} + E_{\rm int } >0 $ with 
$0 < \mu E_{\rm kin} \ll 1$ and $\mu |E_{\rm int }|\ll 1 $, regardless of whether $E_{\rm int }$ is positive or negative. Thus, one concludes that in the low gravity regime it is possible to have $0 < \mu E_\text{B}= M- \mu Q \ll 1$, if $ 0< \mu E_{\rm kin} + \mu E_{\rm int }+ \mu 
E_\text{B}^\text{grav}\ll 1 $. This explanation relies on the difficulty of unequivocally separating 
the total mass in pieces that are intrinsically related by the non-trivial field configuration $\phi (r)$, and also due to the smallness of each of the energy contributions relative to the energy scale $1/\mu$. Had the $Q-$ball solution absent, then even in the 
test-field approximation all the energy contributions would be identically zero and the above separation would not have much 
sense. Notice that the above arguments are consistent with the numerical analysis of the left panels of Figs.~\ref{fig:mr_soliton_sigma006_rescaled} and \ref{fig:Q_Threepanel_charge} and the left and right panels of Fig.~~\ref{fig:bidnerg} showing $\mu M \ll 1$, $\mu^2 Q\ll 1$ and $|\mu E_\text{B}| \ll 1$ in the weak gravity regime associated with $\tilde \phi_0 \ll 1$.

Still in the weak gravity (and high frequency $\omega/\mu\simeq 1$) regimes, one finds a characteristic structure consisting of two cusps (right panel of Fig.~~\ref{fig:Ebind_C_Q_zoom}, which zooms into the middle panel of the same figure).\footnote{Other authors plot $-\mu E_{\text{B}}$ causing the bifurcation diagram to be upside down, however the cusps do not change \cite{tamaki_gravitating_2011, kusmartsev_gravitational_1991, kleihaus_stable_2012}. } The sequence starts on the dilute ($\tilde\phi_0 \ll 1$) branch (we have included the $M(\tilde\phi_0)$ on the left panel for reference). The first cusp (circle markers) is reached as the binding energy becomes increasingly negative, this region is precisely to be the first stable branch. Next, up until the next turning point (square markers), the diagram reaches a local maximum  with positive negative $-\mu E_\text{B}$;   this branch (the one between circle and square markers) corresponds to the first unstable branch. Beyond this segment,  up to a third cusp (diamond markers) the binding energy becomes negative and stability is recovered as one enter the thin-shell regime throughout. Nonlinear evolutions confirm the stability of this compact branch \cite{ge_hair_2024,marks_long-term_2025}. As $\sigma\to 0$ the extent of the negative binding energy region on this branch increases, and very near the maximum mass there is a narrow interval that also hosts LRs. Moreover, this region is correlated  by the rapid rise in compactness and the characteristic backbend (cf. Figs. \ref{fig:CompactnessLeft_MaximumcompactnessRight},\ref{fig:mr_soliton_sigma006_rescaled})  After the third cusp a further turning  point appears (triangle markers).  However,  we are past the maximum mass, the remaining region is the last unstable branch.\footnote{In this regime one also observes the characteristic spiral behavior in the mass–frequency $M(\omega)$ and charge–frequency $Q(\omega)$ curves (and related projections). }

\begin{figure}
    \centering
    \includegraphics[width=1\linewidth]{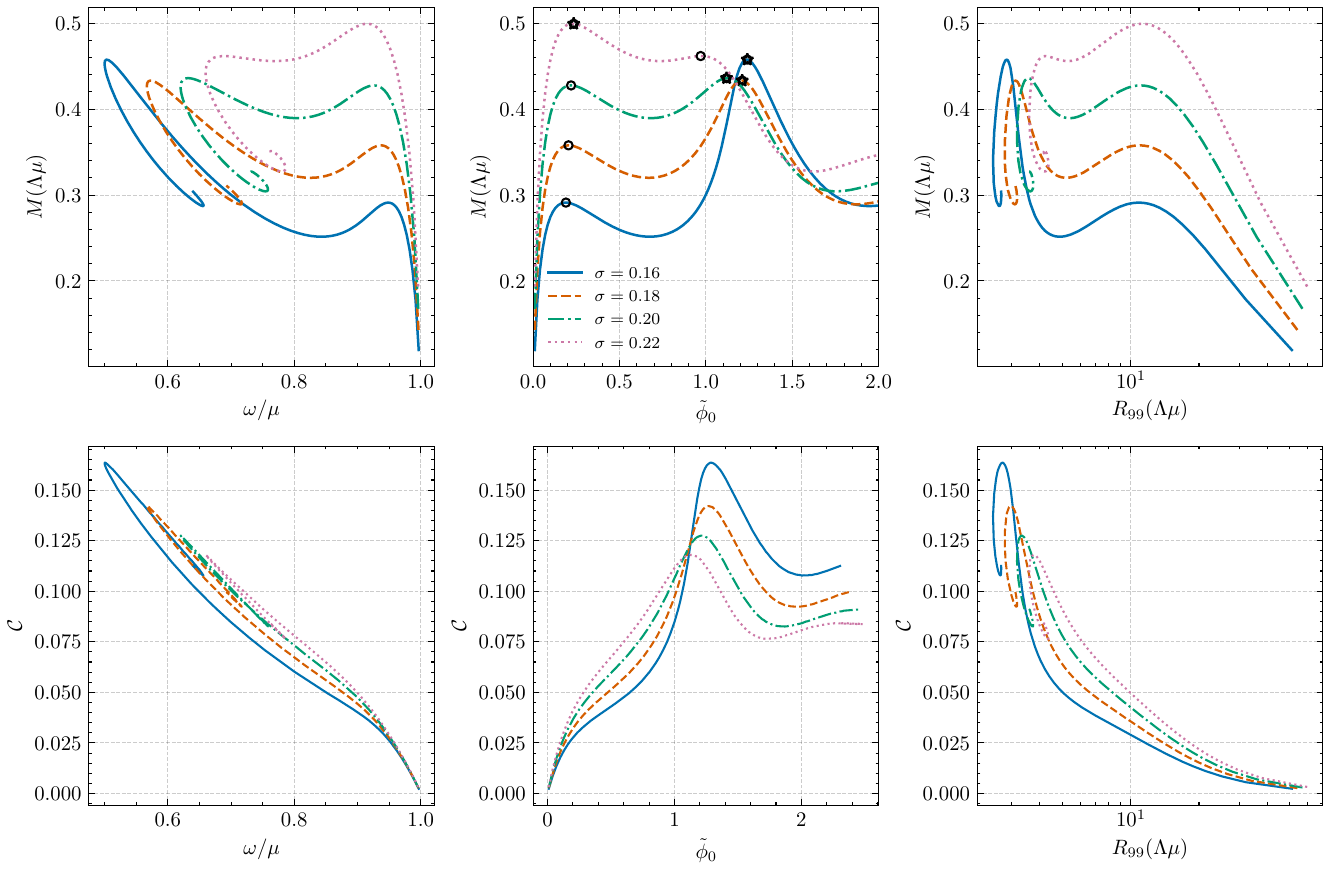}
    \caption{Mass and compactness scaling relations for low–compactness SBSs. Top row: rescaled mass $M(\Lambda\mu)$ as a function of the rescaled frequency $\omega/\mu$ (left), the rescaled central field amplitude $\tilde{\phi}_0$ (middle), and the rescaled effective radius $R_{99}(\Lambda\mu)$ (right).
Bottom row: corresponding compactness $\mathcal{C}$ shown as a function of $\omega/\mu$ (left), $\tilde{\phi}_0$ (middle), and $R_{99}(\Lambda\mu)$ (right).
Each curve corresponds to a fixed value of the parameter $\sigma$. 
}
    \label{fig:Lower_compactness_globalobservations}
\end{figure}

Figure \ref{fig:Lower_compactness_globalobservations} show the results for the less compact SBS 
($\sigma > 0.1$) and can be compared with the more compact SBS ($\sigma < 0.1$) of Figs. \ref{fig:compactness-vs-several} and \ref{fig:mr_soliton_sigma006_rescaled}. We see, for instance, that the maximum (global) mass and maximum compactness are lower than the SBS with lower $\sigma$. Furthermore,  when $\tilde \phi_0 \sim 1$ the transition to the second stable branch becomes smoother. Due to their 
low compactness, in all these cases the SBS do not develop LRs in the stable branch, but barely in the second unstable branch. While interesting, these SBSs are not relevant as BHMs.

To conclude this section it is worth to further elaborate about the stability of ultracompact SBS, in particular, those with a pair of LRs in the second stable branch, and resume the discussion at the end of Sec.~ \ref{sec::Non rotating Boson Stars} within the framework of BHM. 
Let us remind that a paramount condition to consider BHM in terms of horizonless UCOs is {\it stability}. 
Clearly if SBS turn to be unstable just by the very presence of a pair of LR, notably the one for stable circular orbits, then its importance and motivation as BHM would lack any interest as they may 
precisely form a BH under small perturbations. It is then of greatest importance to determine if 
such objects are indeed stable. The analysis of Ref.~~\cite{cunha_light_2017} provides numerical evidence that SBSs are prone to LR instabilities. More specifically, the authors show that the fully nonlinear evolution of a perturbed SBS with a pair of LRs collapses into a rotating BH, leading to the conclusion that SBSs are not viable horizonless UCOs. The authors argue that such instability is generic of all SBS with LR, and thus that other alternative for UCOs as BHM need to be considered. 
Notwithstanding, a subsequent full non-linear analysis in spherical symmetry accompanied with a linear perturbation analysis 
\cite{marks_long-term_2025} led to a different conclusion and showed that such ultracompact SBSs are indeed stable within the same timescales. 
They conclude then that ultracompact SBSs  cannot be ruled out as BHM candidates using stability arguments as in Ref.~~\cite{cunha_exotic_2023}. Additional evidence in this direction is provided by the analysis of the collision of a binary SBS
\cite{siemonsen_nonlinear_2024} where no LR-instability is found in the remnant SBS. Another strong substantiation that the LR-instability conjecture might be false in the rotating case (at least within the same time scales considered in Ref.~\cite{cunha_light_2017})
was recently supplied in Ref.~~\cite{efstafyeva_stability_2025}. That analysis is similar to Ref.~~\cite{cunha_exotic_2023}, except that the authors employ
different codes and tests to proof the robustness of their findings. 

 A final comment in this direction may shed light about this controversy. 
As seen by the panels of the second column of Fig.~\ref{fig::master_profiles} and also from Figures \ref{fig:compactness-vs-several} and \ref{fig:mr_soliton_sigma006_rescaled}, the SBS configurations with LRs are on the second stable branch but very close to the (global) 
maximum mass models. The configurations used as initial data in Refs.~ \cite{cunha_light_2017,marks_long-term_2025,efstafyeva_stability_2025} are very similar to each other and close to the maximum mass models. But perhaps the initial data 
used in Ref.~~\cite{cunha_light_2017} is close but not ``exactly" the same as in Ref.~~\cite{marks_long-term_2025,efstafyeva_stability_2025}) and more close to the maximum compact models, which in principle, are already 
unstable. Other possibility is that both were slightly below the maximum mass model but the methods used to 
perturb their respective initial data are different, perhaps one with larger perturbations than the other so 
that the marginally stable configuration in one case became unstable while the other did not.  
In view of these seemingly opposite results one can 
then conclude that further independent studies along this line are necessary to obtain definite answers about the LR-instability conjecture for SBS.


\section{Incompressible Perfect Fluid Ultracompact Objects (IPFUCO) in GR}
\label{sec:incfluid}
Constant density (incompressible) perfect-fluid (IPF) solutions in Newtonian and relativistic theories of gravity have been widely used in the past to roughly understand 
the origin of some bulk properties of actual astrophysical objects, like the sun and neutron stars (see Ref.~ \cite{Hodek2025} 
for a review and updated analysis). 
These solutions are well known to be unrealistic, but provide enough information to understand heuristically the presence of several features of the more realistic problem.
Clearly, the main drawback is that with such a simplistic model 
one cannot understand the actual nature of the underlying matter of such astrophysical objects, as well as other 
properties of the stars, like the spectrum, temperature, chemical composition etc.
In the context of this work, constant density PF stars provide a useful benchmark to discuss UCOs as one is not limited by the properties of the 
equation of state or the fundamental nature of the fields composing the object. Moreover, for a large central pressure compared with the density, 
one can construct objects (IPFUCO) in hydrostatic equilibrium with the highest possible compactness, which is given by the Buchdahl limit in 
GR ${\cal C}= 4/9$. This limit is independent of the equation of state (EoS) of the matter, provided that some general and 
physical-motivated assumptions are considered \cite{buchdahl_general_1959,Wald}. For this reason IPFUCOs have been employed as toy models of more realistic UCOs in studies of photon spheres, LR structure, and wave dynamics \cite{alho_compactness_2022,pani_gravitational-wave_2018,cardoso_light_2014,urbano_gravitational_2019,hod_number_2017}. In this section we will consider static, spherically symmetric spacetime as in previous sections, but unlike Section \ref{sec::Non rotating Boson Stars}, we consider matter described by 
a perfect fluid  described by the energy-momentum tensor  $T_{ab}= (\rho+ p )u_au_b + p g_{ab}$. The field equations in GR 
take the usual form \cite{Wald}:  
\begin{subequations}\label{eq:TOV_GR_geometric}
\begin{align}
m'(r) &= 4\pi r^{2}\rho, \label{eq:TOV_mprime}\\
\nu'(r) &= \frac{2m(r)+8\pi r^{3}p}{r^{2}-2r m}, \label{eq:TOV_nuprime}\\
p'(r) &= -\frac{\bigl(m+4\pi r^{3}p\bigr)\bigl(\rho+p\bigr)}
{r\bigl(r-2m\bigr)}. \label{eq:TOV_Pprime}
\end{align}
\end{subequations}
When supplemented with an equation of state that closes the system, they form the  Tolman--Oppenheimer--Volkoff (TOV) equations. For a star of uniform density $\rho(r)=\rho_0$ (for $r \leq R$) and $\rho(r)=0 $ (for $r > R$), these equations admit 
the following well known (interior) \Sc exact solution for the matter variables \cite{Wald} (where $R$ the stellar radius, defined by $p(R)=0$, and $M$ is the total gravitational mass):
\begin{align}
\label{eq::GR_FLUID}
p(r) &= \rho_0\,
\frac{\sqrt{1-2\mathcal{C}}-\sqrt{1-2\mathcal{C}\,\frac{r^2}{R^2}}}
{3\sqrt{1-2\mathcal{C}\,\frac{r^2}{R^2}}-\sqrt{1-2\mathcal{C}}},  &
p_0 := p(0) &= \rho_0\,
\frac{1-\sqrt{1-2\mathcal{C}}}{3\sqrt{1-2\mathcal{C}}-1},
\qquad
M=\frac{4\pi}{3}\rho_0 R^3,
\qquad
\mathcal{C}\equiv\frac{M}{R}.
\end{align}
The exact interior solution for the metric components can be consulted in Ref.~ \cite{Wald}. The exterior solution is given by the usual vacuum \Sc solution. We have actually computed the interior and exterior 
solution by solving the system \eqref{eq:TOV_mprime}--\eqref{eq:TOV_Pprime} numerically, as it turns out to 
be more useful to explore the parameter space of solutions for different values of the central pressure $p_0$, and also 
because these solutions can be used as a special case for checking similar numerical solutions obtained for the modified gravity scenario 
that we analyze in the next section.

As $\mathcal{C}$ increases, the required central pressure grows without bound and diverges at the Buchdahl limit $\mathcal{C}=4/9$, which therefore represents the maximal compactness achievable by this model in GR. These are very well known results. However, what is  perhaps less known is that as in the SBS scenario (cf. \ref{sec::Light rings (LR)}), LRs are present in this system as we increase the central pressure and thus the compactness. There is a 
critical central pressure $p_0^{c}$ beyond which LRs appear outside the IPFUCO, and 
which can be traced back to Eq.~ \eqref{eq::GR_FLUID} altogether with the analysis of the effective potential 
for null geodesics \eqref{eq:radial_null}.  In the
exterior (vacuum) Schwarzschild region the null effective potential is
$V_{\rm eff}(r)=L^{2}e^{\nu_{\rm ext}(r)}/r^{2}$ with $e^{\nu_{\rm ext}}=1-2M/r$, and the
LR condition \eqref{eq:LR_cond_general_nu_u} yields the familiar result
$r_{\rm LR}=3M$. An exterior LR therefore exists if and only if $r_{\rm LR}> R$, i.e. $\mathcal{C} > 1/3$. Using Eq.~~\eqref{eq::GR_FLUID}, this translates into $p_0^{c}=\rho_0/\sqrt{3}$.

It is useful to remind that neutron stars (NS) with polytropic EoS cannot have LR if we restrict to the condition of subliminal sound speed \cite{cardoso_testing_2019}. The IPFUCO and in general the IPF models have an infinite sound speed, which also points towards the unrealistic 
nature of these objects. On the other hand, realistic NS EoS satisfying causality and thermodynamic stability lead to even tighter limits, $\mathcal{C} \lesssim 0.33$--$0.35$ at the maximum mass, depending on the specific EoS (see \cite{lattimer_neutron_2007} and references therein). 
Thus, while NS can become highly relativistic, their compactness remains below the above limits. One can evade Buchdahl's limit 
by dropping some assumptions and adopting more exotic forms of matter to construct BHM with a compactness in the range 
$4/9< {\cal C} < 1/2$ \cite{cardoso_testing_2019}.

\begin{figure}
    \centering
    \includegraphics[width=\linewidth]{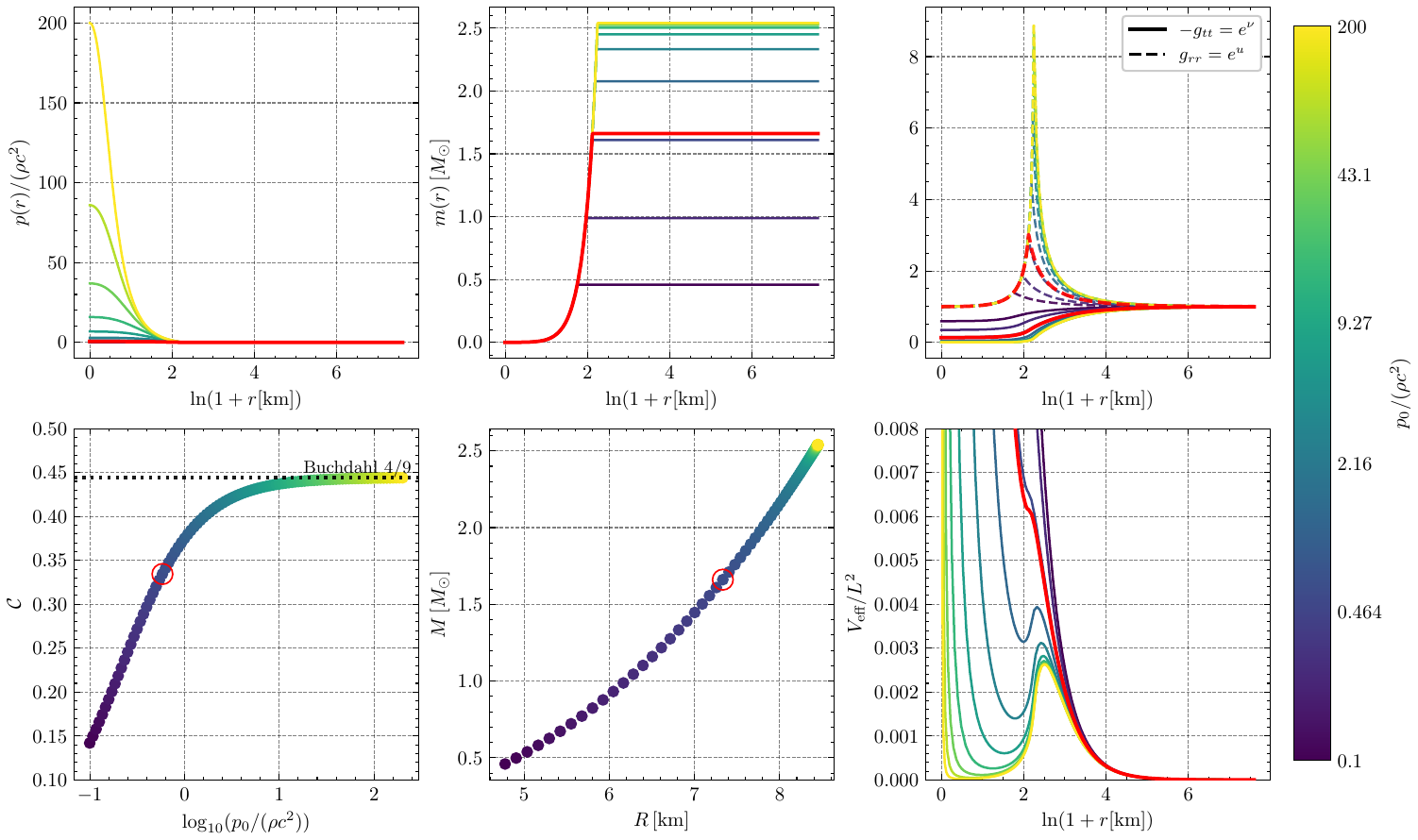}
    \caption{
    Sequence of high compactness solutions for a constant density star with density  $\rho_0=2\times10^{18}[\mathrm{kg}/\mathrm{m^3}]$ in GR. The solutions that start showing LRs are highlighted in red. From top to bottom: 
    pressure profiles $p(r)$ for different values $p_0$ (first row left panel), the mass function $m(r)$ 
    --converging to the total mass $M$-- (first row middle panel), metric components (first row right panel) 
    $-g_{tt}$ (solid lines), $g_{rr}$ (dashed lines); compactness (second row left panel), mass-radius relation (second row middle panel), 
    effective potential for null geodesics $V_\text{eff}$--in units of $L^2$-- (second row right panel).}
    \label{fig:: GRTOV}
\end{figure}

Figure \ref{fig:: GRTOV} depicts some numerical solutions of IPFUCOs that can be compared with the solutions of UCOSBS presented in 
previous section. The models are constructed by fixing a constant density of the order of that one typically found in NS.
The plots show that as the central pressure $p_0$ increases, the compactness follows suit 
(left panel of second row). The critical central pressure $p_0^c=\rho_0/\sqrt{3}$ (indicated with red circles) marks the threshold of solutions below which no LRs are present (right panel of second row).
However, for $p_0 > p_0^c$ we see that the effective potentials develop 
maxima and minima, indicating the presence of unstable and stable LRs, respectively. The red lines in all the panels 
are associated with $p_0^c$. In particular, from the red line for $V_\text{eff}$ (right panel of second row) 
 we appreciate the saddle point prior the emergence of both extrema when $p_0 > p_0^c$. We appreciate that such behaviour 
is very similar to that depicted for $V_\text{eff}$ of SBS in the previous section. 
Thus, increasing the central value $\phi_0$ of the SBS amounts to increasing $p_0$ in IPFUCO, and we see that SBS starts 
behaving like an IPFUCO. Notably, the scalar-field profile looks like a step function, exactly as the constant density model 
IPFUCO. The energy-density of SBS behaves similarly prior the scalar-field reaches the {\it thin shell}, where the highest 
gradients develop, and where the energy-density of SBS increases and decreases sharply. As $p_0\rightarrow \infty$ the minimum 
of $V_\text{eff}$ flattens just like the {\it lapse function} $\alpha=\sqrt{-g_{tt}}$ (right panel of first row). At this point the Buchdahl limit 
is basically reached (left panel of first row). We thus conclude that, 
IPFUCOs are very good mimickers of BHM and help to characterize simply and nicely possible BHM candidates.


\newpage
\section{UCOs in $R^2$--gravity}
\label{sec:fR}

It is natural to ask if BHM as horizonless UCO can be constructed within the framework of relativistic theories of gravity other than GR and, if that is the case, whether the compactness of such objects can be higher or not than the Buchdahl limit in GR. 
In this section we explore such questions within the framework of a simple $f(R)$ metric theory of gravity, which consists of a 
quadratic model similar to the Starobinsky model used for inflation. Moreover, we restrict the analysis of simple IPFUCO within 
this theory, and leave the construction of other kind of UCO --like SBS-- within this or other $f(R)$ gravity theory, for a 
future investigation.

First, we summarize the basic elements of $f(R)$ gravity. As mentioned above, we adopt the \emph{metric} variational principle, as opposed to the Palatini formulation, in which the metric and the connection are treated as independent
variables \cite{sotiriou_fr_2010}. While both approaches reduce to the same
field equations in GR, they generically lead to inequivalent dynamics once the
Lagrangian is promoted to a nonlinear function $f(R)$. We also follow the strategy
used in \cite{jaime_robust_2011,Hodek2025}, which keeps the theory in its
original Jordan-frame form and avoids recasting it into a scalar--tensor theory in the
Einstein frame via a conformal transformation. 

 The general action for an $f(R)$ theory of gravity in a 4-dimensional manifold   is given by 
\begin{equation}
\label{eq::fR_formalism_action_1}
 S[g_{ab}, \Psi] = \int \frac{f(R)}{2\kappa} \sqrt{-g} d ^4x+ S_{M}[g_{ab},\Psi],
\end{equation}
where $\kappa = 8\pi G/c^4$ and $\Psi$ represents collectively and schematically the matter fields. Varying this action
with respect to the metric yields the modified Einstein equations,
\begin{equation}
\label{eq::fR_formalism_1}
f_R R_{ab}-\frac{1}{2}f\, g_{ab}
-\left(\nabla_a\nabla_b-g_{ab}\Box\right)f_R
=\kappa\,T_{ab},
\end{equation}
with $f_R\equiv \partial f/\partial R$, $\Box\equiv g^{ab}\nabla_a\nabla_b$, and
$T_{ab}$ the matter energy--momentum tensor. In the applications considered below, we take $T_{ab}$ to be that of a perfect fluid.

Taking the trace of Eq.~~\eqref{eq::fR_formalism_1} provides an evolution equation for the
Ricci scalar,
\begin{equation}
\label{eq::fR_formalism_3}
\Box R=
\frac{1}{3 f_{RR}}\Big[\kappa T
-3 f_{RRR}(\nabla R)^2
+2f-R f_R\Big],
\end{equation}
where $T\equiv T^{a}{}_{a}$, $(\nabla R)^2\equiv g^{ab}(\nabla_a R)(\nabla_b R)$,
$f_{RR}\equiv \partial_R^2 f$ and $f_{RRR}\equiv \partial_R^3 f$. Substituting
Eq.~~\eqref{eq::fR_formalism_3} back into Eq.~~\eqref{eq::fR_formalism_1} one can rewrite
the system in an Einstein-like form,
\begin{equation}
\label{eq::fR_formalism_4}
G_{ab}=
\frac{1}{f_R}\!\left[
f_{RR}\nabla_a\nabla_b R
+f_{RRR}(\nabla_a R)(\nabla_b R)
-\frac{g_{ab}}{6}\left(R f_R+f+2\kappa T\right)
+\kappa T_{ab}
\right],
\end{equation}
where $G_{ab}=R_{ab}-\tfrac12 g_{ab}R$.

Equations \eqref{eq::fR_formalism_4} and \eqref{eq::fR_formalism_3} constitute the
fundamental field equations in the formulation used here: instead of dealing with a
single fourth-order tensor equation for $g_{ab}$, the dynamics can be expressed as two
coupled second-order PDEs, one for the metric and one for the scalar curvature. This
makes manifest the presence of an additional (scalar) propagating degree of freedom,
carried by $R$ itself. In contrast with GR, where the trace gives the algebraic relation
$R=-\kappa T$ (without cosmological constant $\lambda$), metric $f(R)$ gravity replaces this by a differential relationship 
 \eqref{eq::fR_formalism_3}. The GR limit (with a cosmological constant) is recovered only for
$f(R)=R-2\lambda$, in which case $f_{RR}=0$ and the extra scalar mode disappears.

Like in Sec.~\ref{sec::Non rotating Boson Stars} we restrict the attention to static, spherically symmetric (SSS) spacetimes described by the metric \eqref{eq:Lineelement}, except that we adopt the following abridged notation:
\begin{equation}
\label{eq::f_R_Spherically_line_element}
ds^2 =-n(r)\,dt^2+m(r)\,dr^2+r^2 d\Omega^2,
\end{equation}
where $d\Omega^2=d\theta^2+\sin^2\theta\,d\varphi^2$. Under this ansatz, the coupled
system \eqref{eq::fR_formalism_4}--\eqref{eq::fR_formalism_3} reduces to a set of
ordinary differential equations for $(n,m,R)$ sourced by the matter stress-energy.
A convenient form of the SSS field equations is
\begin{align}
  \label{eq:mprime_1}
   R''= &\frac{1}{3f_{RR}}\left[m(\kappa T+ 2f-Rf_R)-3f_{RRR}R'^2\right] + \left(\frac{m'}{2m}-\frac{n'}{2n}-\frac{2}{r}\right)R',\\
     m'= & \frac{m}{r(2f_{R} + rR'f_{RR})} \bigg\lbrace 2f_{R} (1-m)- 2 m r^2\kappa T^t_t  + \frac{mr^2}{3} \left(Rf_{R}+ f+2\kappa T\right) \\\notag
     & + \frac{rR'f_{RR}}{f_{R}}\left[\frac{mr^2}{3}(2Rf_{R}-f+ \kappa T) - \kappa m r^2(T^t_t+T^r_r)+2(1-m)f_{R}+ 2rR'f_{RR}\right]\bigg\rbrace, \\
      \label{eq:nprime_1_1}
     n' =& \frac{n}{r(2f_{R} + rR'f_{RR})} \left[mr^2(f-Rf_{R}+ 2\kappa T^r_r)+ 2f_{R}(m-1)- 4 rR'f_{RR}\right].
\end{align}
As in GR, the matter sector remains minimally coupled to the metric, so the covariant
conservation law $\nabla_aT^{ab}=0$ is unmodified. For a perfect fluid, this yields a
TOV-like equilibrium condition,
\begin{equation}
\label{eq:fR_TOV_like}
p'=-\frac{\rho+p}{2}\,\frac{n'}{n}.
\end{equation}
Equations \eqref{eq:mprime_1}--\eqref{eq:nprime_1_1}, together with
Eq.~~\eqref{eq:fR_TOV_like} and an equation of state (EoS), define the stellar structure
problem in metric $f(R)$ gravity. As for the general form of the equations, for GR $f(R)=R$, this set of equations reduces to 
the TOV system of hydrostatic equilibrium presented in the previous section.

We specialize to the quadratic (Starobinsky) model \cite{starobinsky_new_1980},
\begin{equation}
\label{eq::f_R_RSquered_model}
f(R)=R+aR^2,\qquad
f_R=1+2aR,\qquad
f_{RR}=2a>0,
\end{equation}
which introduces a single additional scalar degree of freedom whose (Jordan-frame)
mass scale is $m_s^2=\frac{1}{6a}$. Historically, this model was proposed by Starobinsky as a mechanism for inflation in the
early Universe \cite{starobinsky_new_1980}. We choose to measure the parameter $a$ in units of $r_g^2$, where $r_g= GM_\odot/c^2\approx1.48 \text{ km}$ (see \cite{Hodek2025} for details). In Ref.~\cite{Hodek2025} we have reported several neutron star models in this theory by 
exploring possible values for the parameter $a$, and review several related findings by other groups. 
In particular, it has been argued that relativistic-star models can become unstable if $a\gtrsim 2.4 \times 10^{8}\text{cm}^2$ 
\cite{Pretel2020}.

In order to isolate the impact of the modified gravity sector on relativistic stellar structure, we restrict our attention in this work to \emph{constant-density} (incompressible) configurations, $\rho(r)=\rho_0$ inside the object, like in the previous section. As in GR, constant-density stars provide a useful theoretical laboratory: they admit a particularly transparent interpretation of compactness and allow one to probe how close equilibrium solutions can approach ultracompact regimes under the chosen theory assumptions. For each model and parameter choice, we construct sequences labeled, e.g., by the central pressure or by the stellar radius, and compute global observables from the asymptotic behaviour of the metric components $n,m$. Unlike \cite{Hodek2025} we restrict ourselves to 
scenarios where the compactness is high so that the object admits LRs.

We define compactness in the standard way by
\begin{equation}
\label{eq:compactness_def}
\mathcal{C}\equiv \frac{G M}{R_\star c^2},
\end{equation}
where $M$ is the gravitational mass inferred from the asymptotic metric (as computed for the SBS scenario) and $R_\star$ denotes
the stellar radius (defined as the location where $p(R_\star)=0$). Our primary objective is to use these constant-density configurations as a controlled setting to assess the compactness in metric $f(R)$ gravity relative to GR. In particular, we ask whether the additional scalar curvature degree of freedom in the quadratic model can support equilibrium stars with $\mathcal{C}$ exceeding the maximum compactness attainable by incompressible fluid spheres in GR under the same assumptions. For instance, this can happen if the compactness in $R^2$--gravity grows faster 
than the GR value ${C}_{GR}=4\pi \rho R_\star^ 2/3$ for the IPFUCOS, as we increase the density and the central pressure. This is 
because, unlike GR, in $R^2$--gravity  $M\neq 4\pi \rho R_\star^3/3 $ but 
rather $M=4\pi \int_0^\infty \rho_{eff}(r) dr$, where  $\rho_{eff}(r)$ is an effective energy density that includes not only the constant density contribution of the perfect fluid $\rho$ inside the star, but also the effective density contributions associated with the $f(R)$ theory itself which extend beyond the compact support of the fluid (cf. the r.h.s. of Eq.~\eqref{eq::fR_formalism_4}). So, if when increasing the density and pressure the mass $M$ increases more than the radius $R_\star$ does, as compared with GR, it is possible that 
the maximum compactness in $R^2$--gravity exceeds the Buchdahl 
limit, i.e., $4/9 < \mathcal{C}_{R^2,p_\infty}$. 
We show, however, evidence that the opposite happens, i.e., 
$ \mathcal{C}_{R^2,p_\infty} <  4/9$. In fact a similar 
phenomenon occurs in NS models, where the 
maximum compactness in GR turns out to be larger or very similar to that of $R^2$--gravity (depending on the EoS) \cite{Hodek2025}. 

\begin{figure}
    \centering
    \includegraphics[width=1\linewidth]{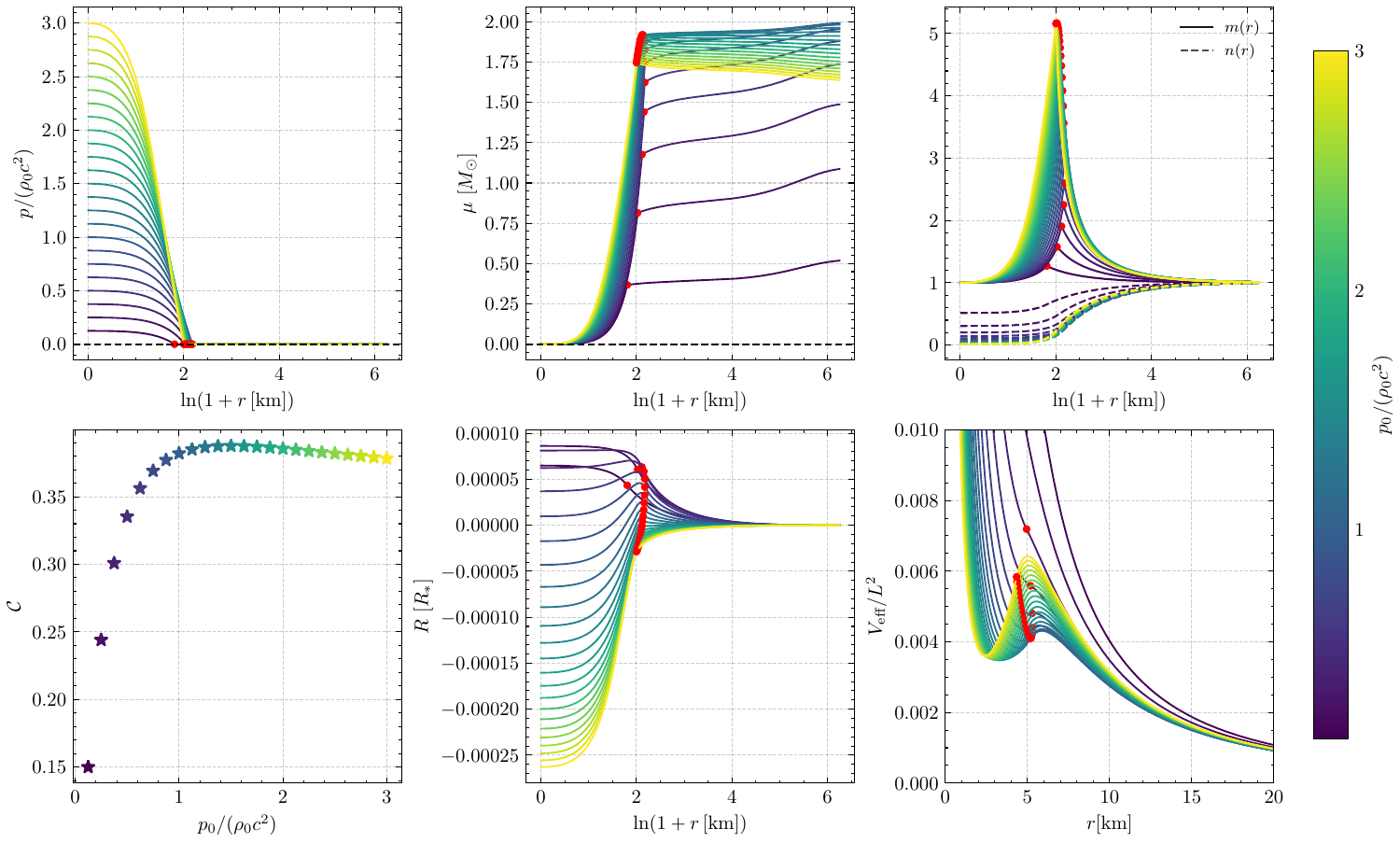}
    \caption{Profiles for constant–density stars for stellar sequence  in $R^2$--gravity at fixed $a/r_g^2=10^3$ and fixed $\rho_0$ (here $\rho_0=2\times10^{18}[\mathrm{kg}/\mathrm{m^3}]$). Each curve corresponds to one solution labeled by its central pressure; color encodes the scaled pressure $p_0/(\rho_0 c^2)$. Top row: (left) pressure profile $p/(\rho_0 c^2)$ vs. radius, (center) enclosed gravitational mass $\mu(r)$ in $M_\odot$ units, (right) metric functions $m(r)$ (solid) and $n(r)$ (dashed). Bottom row: (left) compactness $\mathcal{C}=GM/(R_\star c^2)$ vs.\ central pressure $p_0/(\rho_0 c^2)$; the sequence reaches a maximum $\mathcal{C}$ and then decreases at higher $p_0$, (center) Ricci scalar profile $R(r)$ in units of $R_*=r_g^{-2}$, (right) effective potential $V_\text{eff}/L^2=n(r)/r^2$.  Red dots indicates stellar surface $R_\star$ where the pressure vanishes.}
    \label{fig:RSquared_1}
\end{figure}

\begin{figure}
    \centering
    \includegraphics[width=\linewidth]{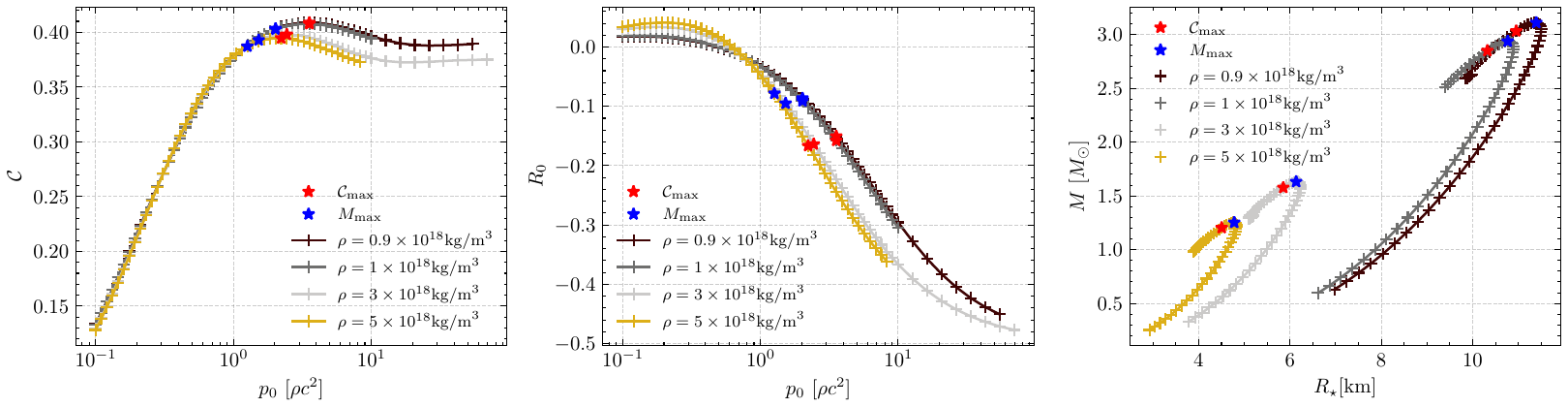}
    \caption{Summary of constant–density stellar solutions in $R^2$--gravity, with $a/r_{g}^{2}=1$. {\it Left:} compactness $\mathcal{C}$ versus the central pressure $p_{0}$. {\it Middle:} Ricci scalar $R(0)=R_{0}$ versus $p_{0}$. {\it Right:} mass–radius relation, $M$ {\it vs.} $R_\star$. We show results for three distinct densities, $\rho\sim 10^{18}\mathrm{kg/m^{3}}$ . The red star marks the maximum-compact configuration and the blue star marks the maximum-massive configuration for each dataset.}
    \label{fig:placeholder3}
\end{figure}


The numerical details for solving the system \eqref{eq:mprime_1}--\eqref{eq:nprime_1_1} are described in Ref.~ \cite{Hodek2025}. 
In Fig.~~\ref{fig:RSquared_1} we present a handful of solutions where we fix the density of the star and vary the central pressure. 
Notice that the central pressure increases (top row first panel) the central value of the scalar field (bottom row middle panel) 
decreases and becomes negative. Moreover, for larger pressures $-\kappa T<0$ since $-T= \rho c^2  - 3 p $, therefore the total effective energy-density 
$\rho_{eff}(r)$ as defined from the r.h.s of \eqref{eq::fR_formalism_4} contains negative contributions, unlike the GR scenario where 
$\rho_{eff}(r)= \rho c^2$ is given only by the perfect fluid energy-density. These negative contributions produce that the total mass $M$ starts decreasing for a sufficiently large $p_0$ (top row, middle panel). In this case, $M$ is given by the mass function $\mu (\infty)$. Note that this function, instead of increasing monotonically 
with $r$, it decreases for large $p_0$. In particular, it decreases (due to the geometric contributions of the $f(R)$ theory to $\rho_{eff}(r)$) 
after the star radius (indicated with red dots). For lower $p_0$ it is the opposite, and we see that $\mu (r)$ increases with $r$. As a consequence, 
the compactness reaches a maximum (unlike GR, where it only reaches a limit) and then it decreases for large $p_0$. As expected, the effective potential for 
null geodesics (bottom row, third panel) develops LRs beyond a critical $p_0$, like in GR.

Figure \ref{fig:placeholder3} summarizes global quantities of the IPFUCOs for three different values of constant density and for an $R^2$-- model 
with a low constant $a$. In these cases, the total mass also decreases for large $p_0$ (right panel), and so the compactness (left panel). Like in the 
example with $a_{1k}=a/r_g^2 = 10^3$, the central value of the Ricci scalar also becomes negative with $p_0$. However, in these three examples, 
the maximum compactness is larger than the case $a_{1k}$ because they are closer to the GR limit, which is reached for $a/r_g^2\ll 1$.

Given the behaviour of IPFUCOs in this theory, it would be interesting to analyse SBS in this framework as well and see if around the regime of degenerate vacuum ($\tilde \phi_0 = 1$) 
negative contributions to $\rho_{eff}(r)$ also appear, and thus, estimate if the maximum compactness of SBS in $f(R)$ gravity also decreases relative to the SBS in GR. 
Needless to say, this scenario entails more technical difficulties as it would be necessary to implement a {\it double shooting} method, one for solving the 
Ricci scalar equation and another one for solving the eigenvalue problem associated with the scalar boson field. We leave this study for a future investigation.

\section{Conclusions}
Black holes are a definite and astonishing prediction of Einstein's GR. There is a robust, albeit indirect, evidence 
about their existence in the universe under different forms: GW sources, powerful engines at the centers of galaxies, sources of 
characteristic patterns of light due to extreme light bending from accretion disks, etc. One of the standard mechanisms of BH formation is the gravitational collapse of stellar objects, like neutron stars or supernovae. However, the same kind of mechanism 
can lead to the formation of undesirable singularities within the interior of BHs, at least classically, although quantum gravity (QG) 
effects might cure those singularities \cite{Rovelli}. Alternatively, ultracompact objects or UCOs might be a remnant 
as the result of an intricate process involving QG aspects of spacetime 
during gravitational collapse \cite{BHmimickers,Mersini2016}. Notwithstanding, at this stage, it is a difficult task to validate that 
possibility with certainty in the absence of a reliable theory of QG. The absence of singularities in BHs might also cure the {\it information loss} paradox, a conundrum of contemporary physics \cite{BHmimickers,Mersini2016}. Another possibility is that BHs simply do not entirely evaporate and information remains stored at their interior \cite{Chen2015}. Solitonic Boson Stars (SBS) are self-gravitating theoretical objects that can become UCOs for a low self-interacting coupling $\sigma$. These objects are not intended to fully replace the BH idea and solve the above problems as they can collapse themselves into BHs, much as ordinary compact stars. However, they are interesting constructs that allow us to explore different features that are not present in other self-gravitating objects, like LR. Moreover, they are the result of an explicit and simple field theory without appealing to more artificial 
constructs of UCOs, like gravastars \cite{Mazur2023} or other similar objects (e.g. fuzzballs, frozen stars) that require an ad-hoc engineering or more exotic assumptions \cite{BHmimickers}.
In this paper we have explored UCOSBS models with values of $\sigma$ as low as $\sigma=0.02$ which are difficult 
to construct using standard limited-precision numerical algorithms. In our case we implement an arbitrary-precision 
algorithm based on \textsc{Julia} which is limited only by memory allocation. Thus, the resulting models can 
reach a compactness higher than previous computed models, and present features that resemble those of incompressible (constant-density) 
perfect fluid models when the pressure is high (IPFUCOs). It remains to be investigated if SBS hosting LRs 
are actually stable or not \cite{cunha_exotic_2023,marks_long-term_2025,efstafyeva_stability_2025}. 
In order to provide an heuristic characterization of those UCOSBS 
we explored the IPFUCOs by showing the emergence of LRs beyond some critical central pressure. These objects are 
much more easier to construct numerically than the SBS and are not limited by the EoS or stability considerations. 
In order to assess if the maximum compactness $\mathcal{C}= 4/9$ of this IPFUCOs (according to GR) 
can be overcome in the framework of other relativistic theories of gravity, we analyzed 
IPFUCOs in a quadratic $f(R)$ gravity. Nevertheless, we conclude suprisingly that this limit is even lower than 
in GR even for high $a$. This trend seems to be not exclusive of IPFUCOs, as in this theory NS can be less compact 
although more massive than in GR \cite{Hodek2025}. It would be interesting to examine this feature in  SBS built in 
$R^2$--gravity and compare the outcome with the SBS analyzed here within GR, but this study is left
for a future work. 

\section*{Acknowledgments}
This work was partially supported by 
DGAPA-UNAM PAPIIT grants IN105223 and IN103926. M.S. acknowledges support from DGAPA-PASPA sabbatical grant. 
We are indebted to C. Palenzuela for thorough discussions and suggestions and for providing numerical data 
to crosscheck our results on SBS.

\appendix
\section{Numerics}

In this appendix, we briefly expand on the details regarding the numerical integration of the EKG system \eqref{eq:Ein_tt}- \eqref{eq:KG}.

\subsection{Dimensionless system}
\label{sec::Dimensionless}
As already discussed in Section \ref{sec::Boundary conditions and reescaling}, an appropriate choice of rescaling variables allows us to cast the system in a dimensionless form. To make this more explicit, let us introduce the following arbitrary variables 
\begin{equation}
\label{eq::EQ_de}
r=r_\ast \tilde r,\qquad \omega=\omega_\ast \tilde\omega,\qquad \phi_0=\phi_\ast \tilde\phi,\qquad V(\phi^2)=V_\ast \tilde V(\tilde\phi^{2}),
\end{equation}
and define the first order variable $\psi:=\phi'=(\phi_\ast/r_\ast)\tilde\psi$. In terms of this variables the EKG system takes the form
\begin{subequations}
\label{eq::EKG_Dimensionless_form}
\begin{align}
\tilde u' &= \frac{1-e^{u}}{\tilde r}
+ \tilde r\Big[
C_1 \tilde\psi^{2}
+ C_2 e^{u-\nu}\tilde\omega^{2}\tilde\phi^{2}
+ C_3 e^{u}\tilde V(\tilde\phi^{2})
\Big], \\
\tilde\nu' &= \frac{e^{u}-1}{\tilde r}
+ \tilde r\Big[
C_1 \tilde\psi^{2}
+ C_2 e^{u-\nu}\tilde\omega^{2}\tilde\phi^{2}
- C_3 e^{u}\tilde V(\tilde\phi^{2})
\Big], \\
\tilde\phi' &= \tilde\psi, \\
\tilde\psi' &= e^{u}\Big[C_4 \tilde V_x(\tilde\phi^{2}) - C_5 e^{-\nu}\tilde\omega^{2}\Big]\tilde\phi
- \Big(\frac{2}{\tilde r} + \frac{\tilde\nu'-\tilde u'}{2}\Big)\tilde\psi,
\end{align}
\end{subequations}
where $\tilde V_x=d\tilde V/d\tilde\phi^{2}$, $V(\tilde\phi^2) = \tilde\phi^2\,(1 - \tilde\phi^2)^2$ and  primes denote $d/d\tilde r$. The dimensionless coefficients read as
\begin{equation}
C_1=\kappa \phi_\ast^{2},\quad
C_2=\kappa \omega_\ast^{2}\phi_\ast^{2} r_\ast^{2},\quad
C_3=\kappa V_\ast r_\ast^{2},\quad
C_4=\frac{V_\ast r_\ast^{2}}{\phi_\ast^{2}},\quad
C_5=\omega_\ast^{2} r_\ast^{2}.
\end{equation}
Written in the form \eqref{eq::EKG_Dimensionless_form}, the system can be seen as a first-order system of the form $dy^i/dr=\mathcal{F}^i(r,y^i)$, where $y^{i}=(\tilde u, \tilde\nu, \tilde\phi, \tilde\psi)$. It can be used to solve a variety of BS (with other kinds of scalar-field potentials). In particular, in this work, we adopted the rescaling \eqref{eq:Reescale_variables} suitable for the solitonic potential, so that
\begin{equation}
\label{eq::dimensionless_coeff}
\quad\quad\quad r^*=\frac1{\Lambda\,\mu},\quad
\omega^*=\Lambda\,\mu,\quad 
\phi_0^*=\frac{\sigma}{\sqrt2},\quad
V^*=\frac{\mu^2\,\sigma^2}{2},
\end{equation}
involving the following coefficients 
\begin{equation}
C_1  = \frac{\kappa\sigma^2}{2}= \frac{\Lambda^2}{2}, \quad
C_2 =  \frac{\kappa\sigma^2}{2}=\frac{\Lambda^2}{2} , \quad
C_3 = \frac{1}{2}, \quad
C_4  = \frac{1}{\kappa\sigma^2}= \frac{1}{\Lambda^2}, \quad
C_5 = 1 \,.
\end{equation}
From the choice of the rescaled variables, Eq.~ \eqref{eq:dQdr-phys} takes the form 
\begin{equation}
\frac{d\tilde q}{d\tilde r}
= 8 \pi\tilde\omega \tilde r^{2}e^{(u-\nu)/2}\tilde\phi(\tilde r)^{2},
\qquad\text{with}\qquad  q=\frac{\sigma^{2}}{2(\Lambda\mu)^{2}} \tilde q, 
\end{equation}
and the binding energy takes the form:
\begin{equation}
\mu E_\text{B}
=
\frac{\tilde M}{\Lambda}
-\frac{\sigma^{2}}{2\Lambda^{2}}\,\tilde Q \,.
\end{equation}
Finally, we introduce the following dimensionless energy density, as well as the corresponding radial and tangential pressures
\begin{equation}
\rho_*\equiv \frac{\sigma^{2}(\Lambda\mu)^{2}}{2},
\qquad
\tilde\rho\equiv \frac{\rho}{\rho_*},
\quad
\tilde P_{\rm rad}\equiv \frac{P_{\rm rad}}{\rho_*},
\quad
\tilde P_{\rm tan}\equiv \frac{P_{\rm tan}}{\rho_*}.
\end{equation}
So that, using Eq. \eqref{eq::dimensionless_coeff} one finds
\begin{subequations}
\begin{align}
\tilde\rho
&=
e^{-\nu}\,\tilde\omega^{2}\,\tilde\phi^{2}
+e^{-u}\tilde\phi'^{2}
+\frac{1}{\Lambda^{2}}\,\tilde\phi^{2}\left(1-\tilde\phi^{2}\right)^{2},
\\
\tilde P_{\rm rad}
&=
e^{-\nu}\,\tilde\omega^{2}\,\tilde\phi^{2}
+e^{-u}\tilde\phi'^{2}
-\frac{1}{\Lambda^{2}}\,\tilde\phi^{2}\left(1-\tilde\phi^{2}\right)^{2},
\\
\tilde P_{\rm tan}
&=
e^{-\nu}\,\tilde\omega^{2}\,\tilde\phi^{2}
-e^{-u}\tilde\phi'^{2}
-\frac{1}{\Lambda^{2}}\,\tilde\phi^{2}\left(1-\tilde\phi^{2}\right)^{2}.
\end{align}
\end{subequations}

\subsection{Shooting method}
\label{secc:appendix_Shooting_method}

\begin{figure}
    \centering
    \includegraphics[width=\linewidth]{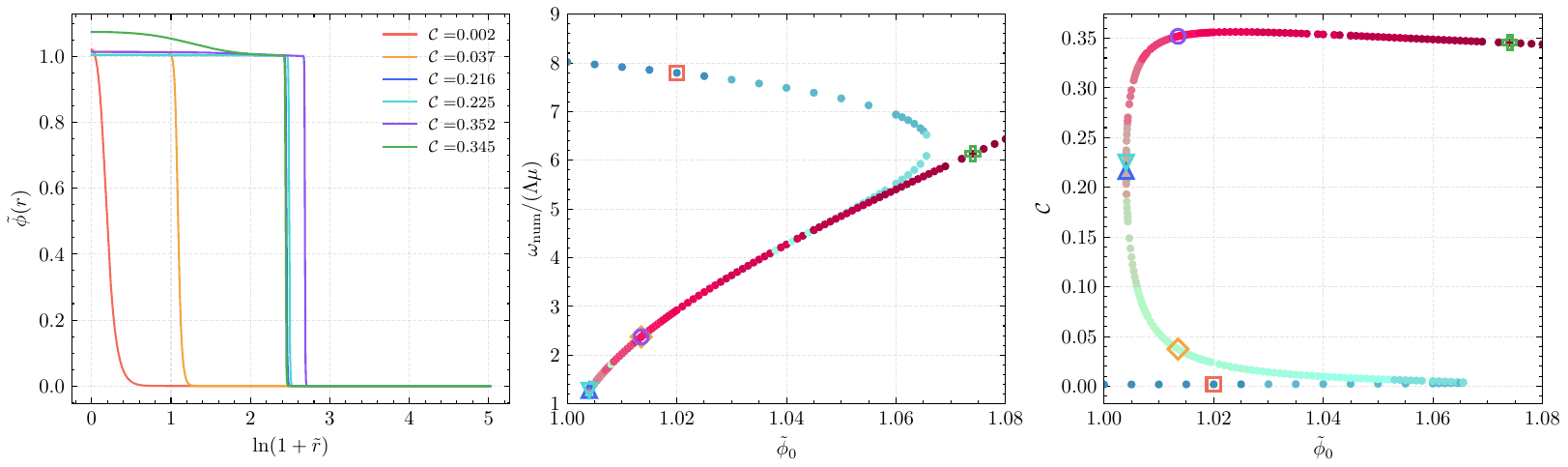}
    \caption{
Radial profiles of six representative SBS configurations for $\sigma=0.02$. 
\textit{Left panel}: scalar field profiles $\tilde\phi(r)$ as a function of $\ln(1+ \tilde r)$ for the six highlighted solutions. 
\textit{Middle panel}: numerical eigenfrequency $\omega_\text{num}/(\Lambda\mu)$ (not rescaled by the lapse) as a function of the central field amplitude $\tilde\phi_0$. Four configurations (triangles, diamond and circle markers) lie in a very narrow $\tilde\phi_0$ interval. 
\textit{Right panel}: compactness $\mathcal{C}$ versus $\tilde\phi_0$, illustrating that configurations with nearly identical frequency can nevertheless exhibit drastically different compactness. 
The six solutions are highlighted with distinct colors and markers consistently across all panels.}
    \label{fig:Numerics_Appendix}
\end{figure}

All equilibrium configurations discussed in this work were obtained by solving the spherically symmetric EKG system
[Eqs.~\eqref{eq:Ein_tt}--\eqref{eq:Ein_thth}] as a boundary-value eigenproblem for the
frequency $\omega_{\text{num}}$. We implemented the code in \textsc{Julia} and adopted a standard outward shooting method: for a chosen central amplitude $\phi_0$ we integrate
the ODE system  \eqref{eq::EKG_Dimensionless_form} from the origin up to a large cutoff $\tilde r_{\max}$ and tune $\omega$ such that the scalar field decays exponentially and
remains nodeless (ground state).

A central numerical difficulty is that the shooting procedure must first identify a
frequency \emph{bracket} $[\omega_{\rm lo},\omega_{\rm hi}]$ such that the corresponding
integrations exhibit opposite \textit{tail behavior} at large radius. Concretely, for fixed
$\phi_0$ an overestimated frequency, typically drives the solution toward an oscillatory
or decaying tail, whereas an underestimated frequency tends to produce a solution that blows up (as the numerical error accumulates).
We therefore monitor the asymptotic behavior of $\phi(\tilde r)$ and its
logarithmic derivative near $\tilde r_{\max}$, and classify each trial integration by  a tail indicator (e.g.\ the sign of $\phi(\tilde r_{\max})\times\phi'(\tilde r_{\max})$,
the number of sign changes). Once a valid bracket is
found, the eigenfrequency is refined by bisection. Special care must be taken for solutions whose eigenfrequency is very close to the one of an excited state (a solution with one or more nodes). In practice, as $\sigma$ decreases and, for each fixed $\sigma$, as we move along the sequence towards larger compactness (precisely where the thin-shell configurations reside), the frequency bracket must be tuned with progressively higher precision. To illustrate this fact, in Fig.~~\ref{fig:Numerics_Appendix} we highlight six distinct configurations, organized in three pairs. Two pairs have eigenfrequencies which differ by only $\Delta\omega_{\text{num}}\sim 10^{-93}$ and $\Delta\omega_{\text{num}}\sim 10^{-20}$ (their respective $\phi_0$ value coincide). As shown in the right panel, the lower pair (triangle markers) have very similar compactness, whereas the upper pair (diamond and circle markers)  exhibits a dramatic difference in $\mathcal{C}$, despite the relative numerical frequency difference.  The middle panel shows $\omega_{\rm num}/(\Lambda\mu)$ as a function of $\tilde\phi_0$ and it exhibits two branches that approach each other as $\tilde\phi_0 \to 1$. The configurations along these branches were the most difficult to find with the shooting algorithm, since it was particularly hard to identify a reliable frequency bracket. 

To cope with this level of precision, we employ arbitrary-precision arithmetic for both the shooting procedure and the ODE integration, increasing the working precision as $\sigma$ decreases or as the desired compactness grows. For the numerical integration of the initial-value problem at each trial frequency $\omega$ we primarily use the implicit, adaptive stiff solvers \texttt{Rodas5()} and \texttt{KenCarp58()} from \texttt{DifferentialEquations.jl}. The method \texttt{Rodas5()} is a 5th-order Rosenbrock--Wanner (linearly implicit) scheme, whereas \texttt{KenCarp58()} is an IMEX Runge–Kutta method of order 5(4) with embedded error control and stiffly accurate implicit stages \cite{kennedy_low-storage_2000, steinebach_construction_2023}. Both algorithms are well-suited to the thin-shell regime, where the presence of sharp gradients and widely separated length scales demands robust stiff integration with tight tolerances.

The price to pay for this robustness is computational cost. The combination of extremely fine bracketing in $\omega_{\text{num}}$, arbitrary-precision arithmetic, and stiff ODE solvers leads to a large number of trial integrations for each data point in the sequences presented in this work. All computations were carried out on a personal laptop, and the most compact solitonic configurations required many hours of cumulative CPU time to obtain convergent solutions with the desired accuracy.

\section{$Q$-balls}
\label{sec:Qball}

In this Appendix we provide a brief overview of the $Q$-ball construction for a single complex scalar
field with a global $U(1)$ symmetry in flat spacetime. The relation between SBS and $Q$-balls is well established in literature. We refer to more detailed treatments on the subject \cite{nugaev_review_2020, heeck_understanding_2021,coleman_q-balls_1985,tamaki_how_2011, volkov_spinning_2002, multamaki_limits_2002}. In particular, the explicit $Q$-ball solution corresponding to the SBS potential \eqref{eq::SBS_Potential} has already been analyzed by \citet{boskovic_soliton_2022}. 

$Q$-balls can be understood as non-topological solitons that arise in the absence of gravitational effects. As explained in the main text, the scalar potential must include terms up to sixth order in the field in order to support $Q$-ball solutions. This requirement contrasts with the quadratic (MBS) and quartic (BS with self-interactions) potentials, which do not admit non-trivial solitonic type solutions. 

\begin{figure}
        \centering
        \includegraphics[width=1\linewidth]{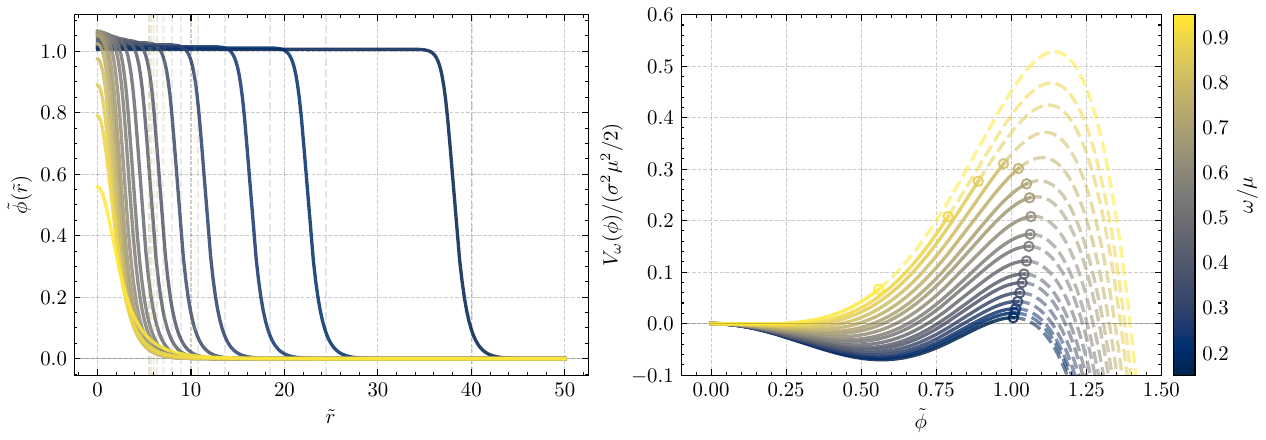}
        \caption{ \textit{Left panel:} scalar-field amplitude $\tilde{\phi}(\tilde r)$ as a function of the rescaled radius, with vertical dashed lines marking $R_{99}$ (the radius enclosing $99\%$ of the total energy). Curves are color-coded by the frequency $\omega/\mu$. \textit{Right panel:} effective potential $V_{\omega}(\phi)$ defined in Eq.~\eqref{eq::Potential_Qballs_effective}; for each configuration, the solid segment corresponds to the field range actually attained along the solution $\tilde{\phi}(\tilde r)$, while the dashed continuation extends $V_{\omega}$ to a common $\tilde{\phi}$-range for visual comparison. The marker indicates  $V_\omega(\tilde{\phi}_{0})$. Similar representations of the effective potential appear in Refs.~~\cite{boskovic_soliton_2022, heeck_understanding_2021}.}
        \label{fig:Q_Ball_profiles}
\end{figure}

Consider the flat spacetime limit of the EKG system \eqref{eq::EKG_Full_system}. Setting
the metric functions to their Minkowski values, $u=\nu=0$, the field equation reduces to a single nonlinear radial ODE for the scalar field amplitude,
\begin{align}
 \label{eq::q_ball_scalar}
  \phi'' + \frac{2}{r}\phi'=\left(\frac{dV(\phi^2)}{d\phi^2} -\omega^2) \right)\phi.
\end{align}
Recall that we are assuming the harmonic dependence on the complex scalar field $\Phi(t,r)= e^{i\omega t}\,\phi(r)$, where $\omega$ is the scalar field frequency. Eq.~ \eqref{eq::q_ball_scalar} can equivalently be derived by varying the action for a complex scalar field in flat spacetime. Following the discussion in \cite{volkov_spinning_2002,kleihaus_rotating_2005} (see also Section II of \cite{zhou_non-topological_2025, boskovic_soliton_2022}), notice that written in this form, Eq.~ \eqref{eq::q_ball_scalar} can be interpreted as a classical particle with position $\phi(r)$ under a ``force" 
given in terms of the scalar-field potential (the r.h.s of \eqref{eq::q_ball_scalar}) and 
endowed with radial (time-dependent) friction term $(2/r)\phi'$. In fact, we can recast it in a more insightful way by means of an effective potential $V_\omega$, namely,  
\begin{equation}
\label{eq::Potential_Qballs_effective}
   \phi '' + 2\phi'/r=-\frac{dV_\omega(\phi)}{d\phi}, \qquad V_\omega=\frac{1}{2}\left(\omega^2\phi^2-U(\phi)\right),
\end{equation}
where we have introduced $U(\phi)\equiv V(\phi^{2})$ (so that $U'(\phi)=dU/d\phi=2\phi dV(\phi^{2})/d\phi^2$). In this language, $Q$-ball solutions correspond to trajectories that start at $\phi(0)=\phi_{0}$ with $\phi'(0)=0$  and asymptotically approach
$\phi(\infty)=0$ as $r\to\infty$, while overcoming the friction effects (see right panel of Fig.~~\ref{fig:Q_Ball_profiles}). Formally, this translates into a set of conditions on the scalar potential which  in turn allows us to determine the frequency range in which $Q$-balls can exist. \citet{coleman_q-balls_1985} showed that a  necessary condition for a nontrivial solution that interpolates between
$\phi_{0}\neq 0$ and $\phi=0$ (at infinity) is that $V_{\omega}$ develops a region where it is
positive away from the origin (a hill), i.e. there exists $\phi_*\neq 0$ such that $U(\phi)/\phi^{2}<\omega^{2}$ attains a non-trivial minimum. On the other hand,  the upper bound follows from the requirement of exponential decay in the
linearized asymptotic region.  Altogether, this leads to
\begin{equation}
\label{eq:omega_bounds}
\omega_{\min}^{2}=\min_{\phi>0}\frac{U(\phi)}{\phi^{2}}
=\frac{U(\phi_{\ast})}{\phi_{\ast}^{2}},\qquad \omega_{\max}^{2}=\frac{1}{2}U''(0),
\end{equation}
so that $Q$-balls exist only for $0<\omega<\mu$. For the potential \eqref{eq::SBS_Potential}.

On the other hand, the energy and the Noether charge of the $Q$-ball are defined, respectively, as:
\begin{align}
\label{eq:Energy_qball_phys}
E
&=\int d^3xT_{00}
=4\pi\int_0^\infty dr\,r^2\Big[\omega^2\phi^2+\phi'^2+V(\phi^2)\Big].\\
Q\label{eq:Charge_qball_phys}
&=\int d^3x\,j^0
=4\pi\int_0^\infty drr^2(2\omega\phi^2)
=8\pi\omega\int_0^\infty drr^2\phi^2.
\end{align}
which are the flat spacetime limit of Eqs. \eqref{eq:mass_adm_def} and \eqref{eq:Q-def}, respectively. For numerical convenience (similar to Sec.~ \ref{sec::Global quantities}), we consider the differential form  of Eqs. \eqref{eq:Energy_qball_phys} and \eqref{eq:Charge_qball_phys},
\begin{align}
    \frac{dq(r)}{dr}&=8\pi\omega\,r^2\,\phi(r)^2, & Q&=\lim_{r\to\infty}q(r),& \frac{de(r)}{dr}&=4\pi r^2\Big(\omega^2\phi^2+\phi'^2+V(\phi^2)\Big),  &E&=\lim_{r\to\infty}e(r).
\end{align}
respectively, and solve them together with Eq.~ \eqref{eq::q_ball_scalar}. The radius of a $Q$-ball can be defined by  $E(R_{99})=0.99E$.  As in the SBS case,  a particular rescaling choice to Eq.~ \eqref{eq::dimensionless_coeff} renders the equations of motion dimensionless.  Namely, by normalizing directly by the mass scale $\mu$, that is, $\tilde\omega=\omega/\mu$, $\tilde r=\mu r$ and $\tilde\phi= \phi\sqrt{2}/\sigma$,\footnote{Here we do not use the combination $\Lambda\mu$ to rescale the $Q$-ball system, i.e., in the absence of the gravitational sector. } we obtain the following system 
\begin{subequations}
\label{eq:Qball_first_order_system_dimless}
\begin{align}
\tilde\phi' &= \tilde\psi,\\
\tilde\psi' &= -\frac{2}{\tilde r}\tilde\psi
-(\tilde\omega^2-1)\tilde\phi-4\tilde\phi^3+3\tilde\phi^5,\\
\tilde q' &= 8\pi\tilde r^2\,\tilde\phi^2,\\
\tilde e' &= 4\pi\tilde r^{2}\Big(\tilde\omega^{2}\tilde\phi^{2}+\tilde\psi^{2}
+\tilde\phi^{2}(1-\tilde\phi^{2})^{2}\Big),
\end{align}
\end{subequations}
where primes now denote derivatives with respect to $\tilde r$ and the first order variable $\tilde\psi\equiv d\tilde\phi/d\tilde r$,  has been defined. The physical charge and energy can be recovered by $q=\tilde q\sigma^2/(2\mu^{2})$ and $e=\tilde e \sigma^2/(2\mu)$. Note that we have written the scalar field potential  \eqref{eq::SBS_Potential} (and its derivative) explicitly. In order to solve the system \eqref{eq:Qball_first_order_system_dimless} we impose regularity at the origin in addition to boundary conditions at spatial infinity. Thus, we ask for 
\begin{equation}
\label{eq:Qball_BCs_dimless}
\tilde\phi(0)=\tilde\phi_0,\quad
\tilde\psi(0)=0,\quad
\tilde q(0)=0,\quad
\tilde e(0)=0,
\qquad
\tilde\phi(\tilde r)\rightarrow 0,
\end{equation}
and construct a sequence of $Q$-ball profiles by fixing $\tilde\phi_0$ and determine  $\tilde\omega$ (ground-state solutions) by a shooting procedure to satisfy the asymptotic behavior, mirroring the eigenvalue strategy used for the self-gravitating configurations discussed in the main text (cf. Appendix \ref{secc:appendix_Shooting_method}).  

\begin{figure}
    \centering
    \includegraphics[width=1\linewidth]{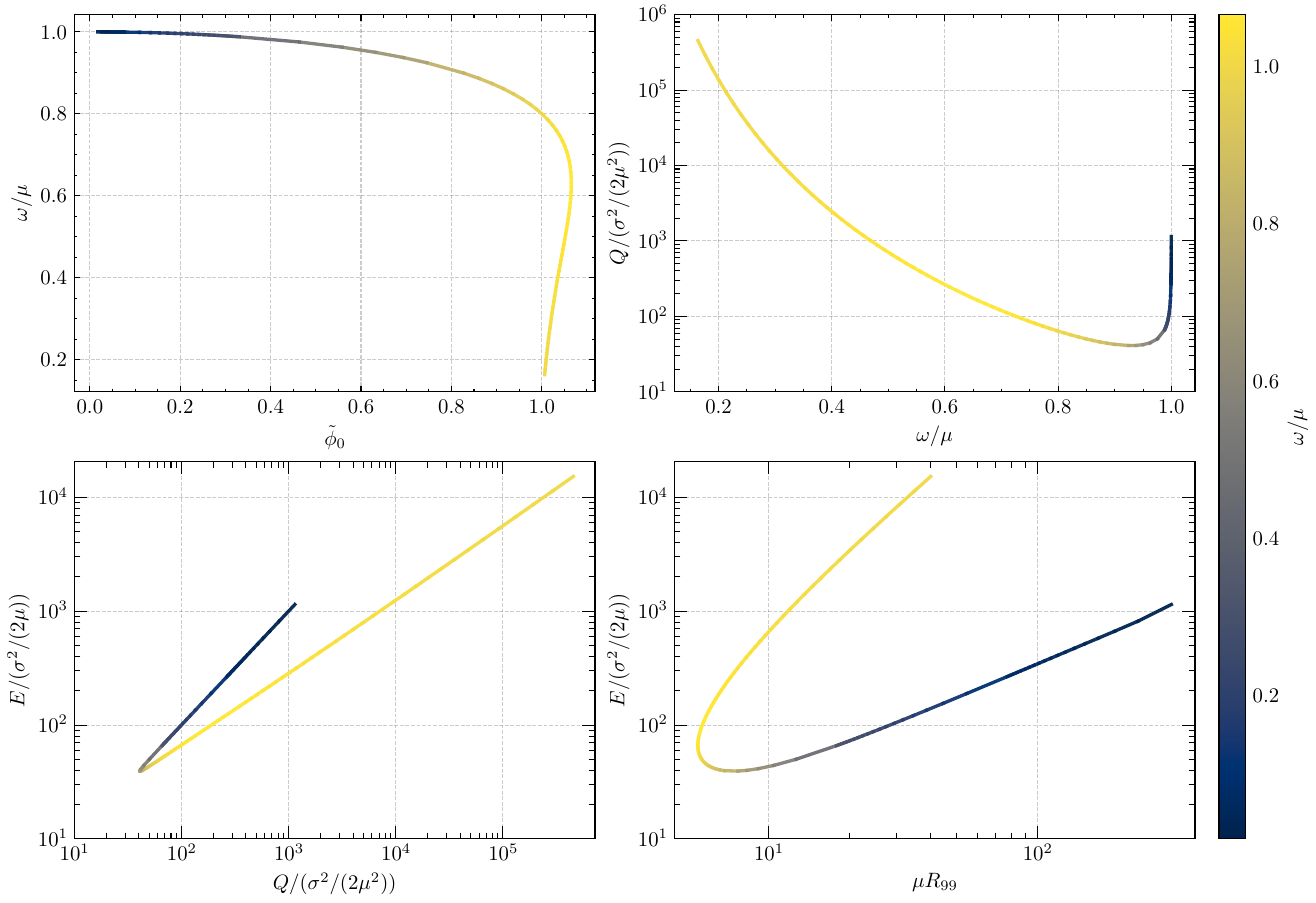}
    \caption{Global properties of the $Q$-ball solution family. From left to right and top to bottom: (i) eigenfrequency $\omega/\mu$ as a function of the central amplitude $\tilde{\phi}_0$; (ii) scalar charge $Q$ versus $\omega/\mu$; (iii) energy $E$ versus charge $Q$; and (iv) energy–radius relation $E$ versus $R_{99}$, where $R_{99}$ encloses $99\%$ of the total energy.}
        \label{fig:Q_Ball_properties}
\end{figure}

In Figure \ref{fig:Q_Ball_profiles} we display 
some examples of numerical solutions for different values $\tilde \phi_0$ that illustrate their general behavior. We observe the characteristic transition (already observed in SBS cf. Fig.~~\ref{fig:sig05}) from solutions in the thick-wall regime  (lower $\omega$, darker colors) to solutions that resemble a step function (thin-shell).\footnote{This is precisely the motivation for Coleman's thin-wall ansatz \cite{coleman_q-balls_1985}.} Observing the effective potential \eqref{eq::Potential_Qballs_effective} in the right panel of the same figure, we see that the solid lines trace the field ``trajectory" along the effective potential for each frequency. For sufficiently low $\omega$, the initial value $\tilde\phi_0$ lies closer to the true vacuum of $V_\omega$. Meanwhile, in Fig.~~\ref{fig:Q_Ball_properties} we show the domain of existence of $Q$-ball family. In contrast to SBS, the existence of $Q$-balls is limited to a range of the scalar field frequency as seen in the top right panel.

Quite interesting is the comparison of the behaviors  $\omega-\tilde\phi_0$ curve with the  $\omega_{\text{num}}-\tilde\phi_0$ of SBS (middle panel of Fig.~~\ref{fig:Numerics_Appendix}). Clearly,  the branch in the $Q$-ball case agrees (qualitatively) with the low compactness branch of the SBS (cf. right panel of 
Fig.~\ref{fig:mr_soliton_sigma006_rescaled}). Figure  \ref{fig:Q_Ball_properties} also exhibits the $E-Q$ curve (bottom left panel) which characterize their (classical) stability \cite{lee_nontopological_1992, zhou_non-topological_2025}. In particular, the stability criterion is that $E< \mu Q$,  and the transition from stability to instability happens very near the cusp of the curve where the solutions experience a turning point (see the $Q-\omega$ curve). 

\appendix

\clearpage
\bibliographystyle{aapmrev4-2.bst}

\bibliography{references.bib} 

\end{document}